\documentclass{LMCS}
\usepackage{enumerate}
\usepackage{hyperref}
\usepackage{amssymb}
\usepackage{amsmath}
\newcommand{\sem}[1]{[\![#1]\!]}
\newcommand{\Terp}[1]{\mathit{Ter(#1)}}
\newcommand{\Ter}{\Terp{\Sigma}}
\newcommand{\Term}[1]{\mathit{Ter}^{#1}(\Sigma)}
\newcommand{\Terinf}{\Term{\infty}}

\newcommand{\Terminfr}[1]{\Terminfp{#1}{\Sigma_R}}

\newcommand{\Terminfp}[2]{\mathit{Ter}^{#1}(#2)}
\newcommand{\emptypos}{\lambda}
\newcommand{\aliencontext}[2]{#1[\![#2]\!]}
\newcommand{\topequr}[1]{\stackrel{#1}{\sim}}

\newcommand{\topequ}[3]{#1\mathrel{\topequr{#2}}#3}

\newcommand{\toplayer}[1]{\lceil #1\rceil}
\newcommand{\ppos}[1]{\mathsf{PPos}(#1)}
\newcommand{\tpos}[1]{\mathsf{TPos}(#1)}
\newcommand{\npos}[1]{\mathsf{Pos}(#1)}
\newcommand{\epos}[1]{\mathsf{Pos}^{m}_{\epsilon}(#1)}
\newcommand{\subterm}[2]{#1/#2}
\newcommand{\Var}{\mathit{Var}}
\newcommand{\weak}{\twoheadrightarrow_w}
\newcommand{\weakweak}{\twoheadrightarrow_{ww}}
\newcommand{\metcom}[1]{#1^{\bullet}}
\newcommand{\proper}{cofinal}

\newcommand{\Foc}{\Focp{f}{\alpha}}
\newcommand{\Focp}[2]{\mathrm{Foc}_{#1,#2}}
\newcommand{\tarsexp}[2]{(#1,\mathcal{O},#2)}
\newcommand{\tars}{\tarsexp{S}{\to}}
\newcommand{\Nat}{\mathcal{N}}

\newcommand{\textiff}{if{f}}

\def\doi{6 (3:18) 2010}
\lmcsheading%
{\doi}
{1--27}
{}
{}
{Sep.~18, 2009}
{Sep.~\phantom06, 2010}
{}

\begin{document}
\title[Modularity of Convergence]{Modularity of Convergence and Strong Convergence in Infinitary Rewriting\rsuper*}
\author{Stefan Kahrs}
\address{University of Kent\\ Department of Computer Science\\ Canterbury CT2 7NF}
\email{S.M.Kahrs@kent.ac.uk}
\titlecomment{{\lsuper*}A shorter version of this paper, only concerned with metric $d_\infty$,
has appeared in the proceedings of RTA 2009 \cite{kahrs-inf-modularity}.}

\begin{abstract}
\noindent Properties of Term Rewriting Systems are called \emph{modular} \textiff{} they are preserved
under (and reflected by) disjoint union, i.e.\ when combining two Term Rewriting Systems with disjoint signatures.
\emph{Convergence} is the property of Infinitary Term Rewriting Systems that all
reduction sequences converge to a limit.  \emph{Strong Convergence}
requires in addition that redex positions in a reduction sequence move arbitrarily deep.

In this paper it is shown that both Convergence and Strong Convergence are
modular properties of non-collapsing Infinitary Term Rewriting Systems, provided
(for convergence) that the term metrics are \emph{granular}. This generalises
known modularity results beyond metric $d_\infty$.
\end{abstract}

\maketitle

\section{Introduction}
Modular properties of Term Rewriting are properties
that are preserved under (and reflected by) the disjoint union of signatures.
They are of particular interest to reason about rewrite systems by \emph{divide and conquer}.

Since Toyama showed that confluence \emph{is} \cite{modular-CR} and
strong normalisation \emph{is not} \cite{Toyama-counter} modular,
the modularity of many further properties have been investigated.
Examples of other modular properties are: uniqueness of normal forms \cite{modular-prop},
weak normalisation, simple termination; examples of non-modular properties
are: completeness, strong confluence.  Modularity proofs for these properties are sometimes
complex \cite{modular-CR}, sometimes simple: the modularity of weak confluence follows directly
from the critical pair lemma \cite{baader-nipkow}.
Many modularity results come with syntactic restrictions, e.g.\ completeness is modular
in the absence of non-left-linear rules \cite{modular-CR-SN-LL} and 
strong normalisation in the absence of collapsing rules \cite{modular-term}.

One aspect of Term Rewriting is that it
can be viewed as a computational model for Functional Programming.
Lazy Functional Programming exhibits a phenomenon that ordinary Term Rewriting does not really capture:
reductions that converge to an infinite result in infinitely many steps.
Partly for that reason, the concept of Infinitary Term Rewriting was devised
in \cite{rewrrewr}; technically, it arises from the finite version
by equipping the signature with a metric $d_m$ and using as its
universe of terms $\Term{m}$, the metric completion of the metric space $(\Ter,d_m)$,
where $\Ter$ is the set of finite terms.

It is common to only consider the metric $d_\infty$
(which goes back to \cite{ArnoldNivat}) in infinitary term rewriting,
ever since its beginning in \cite{rewrrewr};
$d_\infty$ sets the distance $d_\infty(t,t')$ to $2^{-k}$
where $k$ is the length of the shortest position at which the two terms $t$ and $t'$ have different subterm-roots.
This also coincides with a straightforward co-inductive definition of infinitary terms \cite{ketema-thesis}.
The metric completion of $d_\infty$ plays a special role as it is a final co-algebra and thus contains \emph{all}
infinite terms.

However, this is not the only metric of interest: for instance,
the functional programming language Haskell
permits to equip constructors with strictness annotations \cite{haskell98-book}.
These annotations change the universe of terms by ensuring that certain infinite terms
are impossible to construct. For example:
\begin{verbatim}
    data LTree = Null | Bin LTree Int !LTree
\end{verbatim}
The exclamation mark before the third argument of the constructor \texttt{Bin} labels that
argument as strict. For Haskell, this means that to construct a value for
\texttt{Bin t1 n t2} the expression \texttt{t2} needs to be evaluated first.
As a consequence it is impossible to construct an \texttt{LTree} with an
infinite right spine.
The same effect can be achieved by modifying the distance function \cite{kahrs-convergence-acta};
in this case, we could define
\[d_m(\mathtt{Bin}(a,b,c),\mathtt{Bin}(x,y,z))=\max(d_m(c,z),d_m(b,y),\frac{1}{2}\cdot d_m(a,x))\]
With this, the distance between two finite \texttt{LTree} values would be 1 
if their right spines have different lengths. Therefore, all Cauchy-sequences would eventually have
right spines of a fixed finite length with the corresponding implication for metric completion.

Metrics of this form are exactly the \emph{granular} metrics mentioned in title and abstract.
They distinguish between strict and lazy argument positions, which correspond to multiplicative
factors of 1 and $\frac{1}{2}$ (respectively) in the \emph{components} of their associated \emph{ultra-metric maps}.
The associated ultra-metric maps tell us how to compute distances between terms with the same root symbol,
and the components split this task into different argument positions. 
The reasons to treat these metrics separately
are (i) the practical interest in them and (ii) that their restricted form is particularly well-behaved.


Infinitary Term Rewriting views transfinite reductions as reduction sequences
that are also convergent sequences, w.r.t.\ its metric.
In a reduction sequence each term rewrites to its successor.
An infinitary Term Rewriting System (iTRS) is called \emph{convergent} \textiff{} all its reduction sequences converge.

Before \cite{kahrs-inf-modularity} there were few positive results about convergence in the literature,
instead research has focussed on \emph{strong convergence}
which in addition requires that redex positions in a reduction sequence cannot stay short:
if we replace each contracted redex with a random term we would still have a converging sequence.

It is known that strong normalisation is not modular
for finitary Term Rewriting Systems. The well-known counter example from \cite{Toyama-counter}
has a strongly normalising TRS with rule
$F(0,1,x)\to F(x,x,x)$, and another one with rules $G(x,y)\to x$, $G(x,y)\to y$; their combination fails
to be strong normalising as the term $F(0,1,G(0,1))$ reduces to itself in three steps.
As we shall see, for the discrete metric strong convergence and strong normalisation coincide
and therefore modularity of strong convergence can fail for other term metrics than $d_{\infty}$.
The example also shows that ``finite convergence'', the convergence of all reduction sequences starting
from a finite term, fails to be modular in both weak and strong form \cite{simonsen-modularity}.

Simonsen \cite{simonsen-modularity} proved that strong convergence is a modular property
for left-linear iTRSs (for metric $d_\infty$);
the left-linearity condition is indeed unnecessary \cite{kahrs-inf-modularity}.
Proving such results for ordinary convergence is more complicated.
\begin{exa}\label{exnonlin}$R_1=\{ F(x,x,y)\to F(x,y,x) \} \qquad R_2=\{ 0\to S(0)\}$\\
Both systems $R_1$ and $R_2$ are converging.
In the combination we have reductions like
$F(0,0,0)\to F(0,0,1)\to F(0,1,0)\to F(1,1,0)\to F(1,0,1)\to F(1,1,1)$,
with $1=S(0)$.
If we project position $2$ (or $3$) from this reduction sequence,
thus extracting a sequence of elements from the other system then we get a sequence
where the values go down as well as up, i.e.\ on the partial order induced by $R_2$:
$0,0,1,1,0,1,1,\ldots$
For an overall convergence proof we need to demonstrate
why that projected sequence is supposed to be converging, and that is not trivial.
\end{exa}

If a reduction sequence in the disjoint union of two iTRSs
is converging but not strongly converging then its top layer
``remains active'' throughout the sequence and in particular it can rearrange
its principal subterms, as happened with $0$ and $1$ in Example \ref{exnonlin}.
Informally, principal subterms are those subterms where first a change of signature occurs.

These rearrangements are strongly constrained though by the fact
that the iTRS associated with the top layer is itself converging.
In particular, it can be shown that the subterms at a principal position
in the limit form a so-called \emph{focussed sequence};
these have peculiar properties: (i) any infinite subsequence of a focussed sequence
contains an infinite weak reduction sequence; (ii) any non-converging focussed sequence
contains a subsequence which is a non-converging weak reduction sequence.
Here, ``weak reduction sequences'' are reduction sequences with gaps: an element of such a sequence
does not necessarily reduce to its neighbour in a single reduction step, this can take any number of steps.
For  abstract reduction systems over Hausdorff spaces, convergence of all reduction sequences and convergence
of all focussed sequences are equivalent properties (Proposition \ref{focussedprop}).

This implies that any reduction sequence in the combined iTRS must converge
``layer by layer'', and therefore is converging w.r.t.\ the metric $d_\infty$.
Thus such a sequence can only fail to converge if the infinite term
the sequence is converging to (under metric $d_\infty$) does not exist in the metric completion
derived from the combined system's own metric $d_m$.
For example, if one system has the rule $F(x)\to G(x)$ and the other a unary function symbol $H$
then the infinite term $t=F(H(t))$ may exist under the combined metric whilst $u=G(H(u))$ may not,
in which case the combined system would fail to converge. In granular term metrics such rules
can be dismissed as they instantly break convergence; in the example, $a=F(a)$ would and $b=G(b)$
would not exist.

\section{Preliminaries}
To make the terminology of this paper self-contained,
this section contains a reminder of some basic definitions and facts
about topological and metric spaces \cite{gamelin,copson-metric}, as well as sequences over them.

A \emph{topological space} is a pair $(S,\mathcal{O})$ where $S$ is a set and $\mathcal{O}$
a subset of $\wp(S)$ such that it is closed under finite intersections and arbitrary unions, and
$\emptyset,S\in\mathcal{O}$.
The elements of $\mathcal{O}$ are called \emph{open} sets, their complements w.r.t.\ $S$
are called \emph{closed} sets.
A function $f:A\to B$ between topological spaces is called continuous \textiff{} $f^{-1}(X)$ is open whenever $X$ is open.

An \emph{open base} $B$ is a subset of $\wp(S)$; its associated topology $\mathcal{O}$ is the smallest set
such that $B\subseteq\mathcal{O}$ and that $(S,\mathcal{O})$ is a topological space.

A \emph{neighbourhood} of
$x\in S$ is a set $N\subseteq S$ for which there is an $A\in\mathcal{O}$ such that
$x\in A\subseteq N$. A topological space is called \emph{Hausdorff} \textiff{} any two distinct elements of $S$
have disjoint neighbourhoods. It is called \emph{discrete} if $\mathcal{O}=\wp(S)$.

A \emph{sequence} in a topological space $(S,\mathcal{O})$ is a continuous function $f:\alpha\to S$,
where $\alpha$ is an ordinal. The topology on ordinals $\alpha$ is defined as follows:
for every $x<\alpha$ the sets $\{y\mid y<x\}$ and $\{y\mid x<y\}$ are open and form an \emph{open base},
i.e.\ all other open sets arise through finite intersections and arbitrary unions of these sets.
The sequence $f$ is called \emph{open} if $\alpha$ is a limit ordinal, otherwise it is \emph{closed}.

An open sequence $f:\alpha\to S$ \emph{diverges} \textiff{} it cannot be extended to a sequence
$f':\alpha+1\to S$ where $\forall \gamma<\alpha.\:f'(\gamma)=f(\gamma)$. Otherwise it \emph{converges}.
For example, if $S$ is the set $\{0,1\}$ and $f:\omega\to S$ is given as $f(n)=n \mod 2$ then $f$ diverges
for the discrete topology on $S$ ($\mathcal{O}=\wp(S)$), but it converges on the other 3 topologies one can
define on $S$.

A \emph{subsequence} of $f$ is a strictly monotonic and continuous function $g:\beta\to\alpha$.
$g$ is called \emph{\proper{}} \cite{set-hajnal} \textiff{}
$\forall\gamma.\:\gamma<\alpha\Rightarrow\exists\zeta.\zeta<\beta\wedge\:g(\zeta)\geq\gamma$.
We say that a subsequence $g$ converges \textiff{} the sequence $f\circ g$ does.
Clearly, an open sequence converges \textiff{} all its subsequences do.
Note: (i) \proper{} subsequences compose: if $g$ is a \proper{} subsequence of $f$ and $h$ a \proper{}
subsequence of $g$ then $g\circ h$ is a \proper{} subsequence of $f$; (ii) subsequences
are \emph{strictly normal} functions on ordinals \cite{set-potter}, i.e.\ their application
distributes over suprema.

An open sequence $f:\alpha\to S$ \emph{converges to $c$} \textiff{} it stays eventually within any neighbourhood of $c$.
Notice that this notion coincides with a neighbourhood-based definition of being
``continuous at $\alpha$'' which implies that if $f$ converges to $c$ then it converges,
and dually if it converges it must converge to some value.

A \emph{metric space} is a pair $(M,d)$ where $M$ is a set and $d:M\times M\to\mathcal{R}$
is a distance function with the properties $d(x,y)=0\iff x=y$ and $d(x,y)+d(x,z)\geq d(y,z)$
which implies that $d$ is commutative and returns only positive values \cite{copson-metric}.
Metric spaces are topological spaces:
$A\subseteq M$ is open \textiff{} $\forall x\in A.\:\exists\epsilon>0.\:\forall y\in M.\:d(x,y)<\epsilon\Rightarrow y\in A$.
Metric spaces are Hausdorff. A special case of metric spaces are \emph{ultra-metric spaces} which have
a stronger triangle inequality: $\max(d(x,y),d(x,z))\geq d(y,z)$.

The \emph{diameter} of a subset $A\subseteq M$ is the supremum of all distances between elements
of $A$. The \emph{diameter} of an open sequence $f:\alpha\to M$ is the smallest $\epsilon\geq 0$
such that $\forall \beta<\alpha.\:\exists \gamma_1,\gamma_2\geq\beta.\:d(f(\gamma_1),f(\gamma_2))\geq\epsilon$.

A \emph{Cauchy-sequence} over $(M,d)$ is a function $f:\omega\to M$ such that
$\forall\epsilon>0.\:\exists k.\:\forall m,n\geq k.\:d(f(m),f(n))<\epsilon$.
Thus Cauchy-sequences are precisely those with a diameter of 0.
The metric space is called \emph{complete} \textiff{} all its Cauchy-sequences converge;
each metric space $M$ has a completion $\metcom{M}$, which is unique up to isomorphism.
In infinitary term rewriting, metric completion is used to generate infinite terms from finite terms.
A function $f:M\to N$ between metric spaces $(M,d_M)$ and $(N,d_N)$ is called \emph{uniformly continuous}
\textiff{} $\forall\epsilon>0.\:\exists\delta>0.\:\forall x\in M.\:\forall y\in M.\:
d_M(x,y)<\delta\Rightarrow d_N(f(x),f(y))<\epsilon$.
A uniformly continuous function $f:M\to N$ has a unique continuous
extension to $\metcom{M}\to\metcom{N}$. 
Continuous functions preserve converging sequences,
uniformly continuous functions also preserve Cauchy-sequences.
A special case are \emph{non-expansive}
functions, characterised by the property $d_N(f(x),f(y))\leq d_M(x,y)$.
If even $d_N(f(x),f(y))= d_M(x,y)$ then $f$ is an \emph{isometric embedding}. Isometric embeddings both preserve
and reflect Cauchy-sequences, and --- if the spaces are complete --- converging sequences.

\section{Reduction systems and convergence}

A \emph{topological abstract reduction system} (short: TARS) is a structure $(S,\mathcal{O},\to)$ such that
$(S,\mathcal{O})$ is a topological space and $(S,\to)$ an abstract reduction system,
i.e.\ $\to$ is a binary relation on $S$. In the world of infinitary terms the underlying topological
spaces are typically ultra-metric spaces, see \cite[Proposition 3]{kahrs-convergence-acta} and
\cite[Section 4.3]{hlcs:effective-algebra}.
The reason for the added generality of this section is that (i) the notion of convergence is a topological concept,
(ii) certain convergence results about abstract reduction systems
can be proven at this more general level, and (iii) not all forms of infinitary rewriting derive from metric spaces,
e.g.\ for infinitary $\lambda$-calculus B\"ohm-Trees relate to the Scott-topology \cite{lambda}.

A reduction sequence in a TARS $(S,\mathcal{O},\to)$ is a sequence $f:\alpha\to S$ such
that $\forall\beta.\beta+1<\alpha\Rightarrow f(\beta)\to f(\beta+1)$. Recall that being a sequence entails
the continuity of $f$.

For infinitary term rewriting we are specifically interested in TARS where the
underlying topological space is a metric space. This gives rise to the notion
of a \emph{metric abstract reduction systems} (short: MARS) $(M,d,\to)$
where $(M,d)$ is a metric space, and $\to$ a relation on $M$. 

A TARS is called \emph{convergent} \textiff{} all its open reduction sequences converge.
Otherwise we call it \emph{divergent}. 

\begin{prop}\label{divergence}
If the open sequence $f:\alpha\to S$ diverges then for every $s\in S$ there is a neighbourhood
$N_s$, an ordinal $\beta_s$ and a \proper{} subsequence $w_s:\beta_s\to\alpha$ of $f$ such
that $\forall\gamma<\beta_s.\:f(w_s(\gamma))\notin N_s$.
\end{prop}
\begin{proof}
Otherwise, there would be an $s$ (possibly many) for which these premises are not true.
In that case $f$ converges to $s$.
\end{proof}
For topological spaces that are not Hausdorff it is possible that sequences converge to more than one value.
For example, consider $\omega$ with the cofinite topology:
a subset $A\subseteq\omega$ is open \textiff{} either $A$ is empty or $\omega\setminus A$ is finite.
In that topology the identity function on $\omega$ converges to all numbers,
because the sequence $1,2,3,4,\ldots$ will eventually stay within any cofinite set of natural numbers.

\emph{Weak reduction} $\weak$ of a TARS is defined as follows:
$t\weak u$ \textiff{} there exists a closed reduction sequence $f:\alpha+1\to S$ with $f(0)=t$ and
$f(\alpha)=u$. Clearly, the relation $\weak$ is transitive. We also write $\weakweak$ for
the weak reduction of the TARS $(S,\mathcal{O},\weak)$.
Moreover, we call reduction sequences of the TARS $(S,\mathcal{O},\weak)$ \emph{weak reduction sequences}
of the original TARS $(S,\mathcal{O},\to)$. These are the earlier mentioned ``reduction sequences with gaps''.

\begin{prop}\label{weakweak}
Let $(S,\mathcal{O},\to)$ be a convergent TARS, where $(S,\mathcal{O})$ is Hausdorff.
If $t\weakweak u$ then $t\weak u$.
\end{prop}
\begin{proof}
We can prove the result by induction on the indexing ordinal $\alpha$
of the witnessing sequence $f:\alpha+1\to S$ for $t\weakweak u$.
The base case $\alpha=0$ is trivial and if $\alpha$ is a successor ordinal
the result follows by the induction hypothesis and transitivity of $\weak$.

Otherwise, $\alpha$ is a limit ordinal.
For any $\beta<\alpha$ we have $t\weakweak f(\beta)$ and thus by the induction
hypothesis $t\weak f(\beta)$. This reduction has an associated indexing function $g_\beta:\gamma_\beta\to S$.
The increasing sequence of ordinals $\gamma_\beta$ (with $\beta$ approaching $\alpha$)
must converge to some ordinal $\gamma$ \cite[page 290]{sierpinski}.
W.l.o.g.\ we can assume that this is a limit ordinal as
otherwise almost all reductions in the image of $f$ would be empty.
Moreover the functions $g_\beta$ agree on their common domain.
In their limit they thus extend to a function $g:\gamma\to S$ which is an open reduction sequence.
As the original TARS is convergent $g$ can be extended to a closed sequence $g':\gamma+1\to S$.
Since both $f(\alpha)$ and $g'(\gamma)$ are limits of the sequence $f(\beta)$ with $\beta$ approaching $\alpha$
we must have $g'(\gamma)=f(\alpha)$ and the result follows.
\end{proof}
It is easiest to explain the meaning of Proposition \ref{weakweak} by showing how the property fails
if we drop the condition that the TARS is convergent.
Consider the following example, given as an infinitary string rewriting system:
\begin{exa}
$\{ {BE \to CSE},~ {AC \to AB},~ {BS\to SB},~{SC\to CS} \} $\\
In this system we have for any $n$:
\[ABS^nE\to^*AS^nBE\to AS^nCSE\to^*ACS^{n+1}E\to ABS^{n+1}E\]
Thus $ABS^nE\to^*ABS^{n+1}E$ and also $ABS^nE\weak ABS^{n+1}E$.
Starting from $n=0$ we get $ABE\weakweak ABS^\infty$ as the result is the limit 
of all $ABS^{n}E$. However, we do not have $ABE\weak ABS^\infty$ as intermediate
reduction results in the sequences for $ABS^nE\weak ABS^{n+1}E$ will change the $B$ to an $S$.
The example does not contradict the proposition as the system is not converging:
we have $ABS^nE\weak AS^nBE\weak ABS^{n+1}E\weak\ldots$. 
\end{exa}

The requirement that the topological space is Hausdorff is not merely used in (the last part of) the proof,
the property can indeed fail for non-Hausdorff-spaces. 
\begin{exa}\label{nonhaus}
We define a TARS $E$ with carrier set $\omega$
and the following sets as an open base:
\begin{enumerate}[$\bullet$]
\item $S_n=\{ m\mid m\geq n\}$, for any $n\in\omega$
\item the set of all even numbers
\end{enumerate}
Its relation $\to$ is given as $n\to n+1$, for all $n$.
\end{exa}
For the topology of $E$ every element has a smallest neighbourhood, which is subset of all its neighbourhoods.
For an odd number $n$ it is the set $S_n$, for an even number $n$ it is $S_n$ intersected with the set of
all even numbers.

Since $S_1=\omega\setminus\{0\}$, sequences converge to 1, unless they have a cofinal
subsequence that is constantly 0. In that case the sequence might converge to 0, but only if it does not
contain a cofinal subsequence in which all numbers are odd.
As a whole the TARS is convergent as all its reduction sequences converge to 1.

But the proposition about the weak reduction relation
does not hold here: we have $2\weakweak 0$ but not $2\weak 0$.
The reason is that the reduction sequence $2\to 3\to 4\to\cdots$ converges to
all odd numbers, but to no even numbers. By omitting every other element the
sequence gives rise to the weak reduction sequence
$2\weak 4\weak 6\weak\cdots$ which converges to all numbers, including 0;
but there is no reduction sequence starting from 2 that would converge to 0.

Aside: it is possible that $\weak$ and $\weakweak$ coincide
without $\tars$ being convergent --- e.g.\ we can choose ${\to}=S\times S$.

\begin{prop}\label{weakconverge}
Let $\tars$ be a convergent TARS where $(S,\mathcal{O})$ is Hausdorff.
Then $\tarsexp{S}{\weak}$ is also convergent.
\end{prop}
\begin{proof}
Using the same construction as in the proof of Proposition \ref{weakweak}
we can expand an open $\weak$-reduction into an open $\to$-reduction.
The convergence assumption gives us a limit to the latter which must also be a limit to the former.
\end{proof}
As the proof is based on Proposition \ref{weakweak} we still require
that the topology is Hausdorff, and
the property can indeed fail for non-Hausdorff spaces, as Example \ref{nonhaus} provides again a counterexample:
we can define a weak reduction sequence of length $\omega$ as follows:
$0\to 1\to 2\weak 4\weak 6\weak\ldots$ which converges to $0$.
By iterating this reduction we can construct a weak open reduction sequence $f:\omega^2\to\omega$ such that
for all limit ordinals $\lambda<\omega^2$ we have $f(\lambda)=0\wedge f(\lambda+1)=1\wedge f(\lambda+n+2)=2n+2$.
This fails to converge under the topology of $E$, as both $g:\omega\to\omega^2, g(n)=\omega\cdot n$ and
$h:\omega\to\omega^2, h(n)=\omega\cdot n+1$ are cofinal;
$f\circ g$ and $f\circ h$ are constantly 0 and 1, respectively,
and thus miss the smallest neighbourhoods of all other numbers.

\begin{cor}\label{weakred}
A TARS is convergent if and only if all its weak reduction sequences converge, provided its topology is Hausdorff.
\end{cor}
\begin{proof}
If the TARS is not convergent then the result follows because any reduction sequence is also a weak reduction sequence.
Otherwise Proposition \ref{weakconverge} applies.
\end{proof}
Given a TARS $\tars$ and a sequence $f:\alpha\to S$,
the predicate $\Foc$ on $\alpha$ is defined as follows:
\[ \Foc(\beta)\iff \forall\gamma.\:\alpha>\gamma\geq\beta\Rightarrow\exists\zeta.\:\alpha>\zeta
\wedge\forall\kappa.\:\alpha>\kappa\geq\zeta\Rightarrow f(\gamma)\weak f(\kappa)\]
We have: (i) if $\Foc(\beta)$ then $\Foc(\gamma)$ for all $\gamma\geq\beta$;
(ii) if $\Foc(\beta)$ then $\exists\gamma>\beta.\:f(\beta)\weak f(\gamma)$.
The sequence $f$ is called \emph{focussed} \textiff{} $\exists\beta.\:\Foc(\beta)$.

In words: a sequence is focussed \textiff{} every of its elements sufficiently close to $\alpha$
weakly reduces to all its elements sufficiently close to $\alpha$.
In particular, all weak reduction sequences are focussed, but not vice versa. 

Of interest are specifically \emph{open} focussed sequences. 
Closed sequences ($\alpha=\alpha'+1$) are trivially focussed: to prove $\exists\beta.\Foc(\beta)$ we can choose
$\beta=\alpha'$ which determines $\gamma=\alpha'$, and $\zeta$ can also be chosen as $\alpha'$
which in turn determines $\kappa=\alpha'$, and $f(\alpha')\weak f(\alpha')$ holds by reflexivity of $\weak$.

\begin{prop}\label{focussedsub}
Every \proper{} subsequence $g$ of a focussed sequence $f$ has itself
a \proper{} subsequence $h$ such that $f\circ g\circ h$ is a \proper{} weak reduction sequence.
\end{prop}
\begin{proof}
Proposition 4 in \cite{kahrs-inf-modularity} claimed this for MARS, but the proof works unchanged
for TARS, with no constraints on the topology.
\end{proof}


\begin{prop}\label{focussedprop}
A TARS converges \textiff{} all its focussed sequences converge, provided the topology is Hausdorff.
\end{prop}
\begin{proof}
Since all weak reduction sequences are focussed the $\Leftarrow$ implication follows from Corollary
\ref{weakred}.
Now suppose $f:\alpha\to S$ is a divergent focussed sequence; from it,
we need to construct a divergent reduction sequence,
and by Corollary \ref{weakred} it suffices if it is a weak reduction sequence.

Take any element $c\in S$. Proposition \ref{divergence} gives us
a \proper{} subsequence $w_c:\beta_c\to\alpha$ of $f$ witnessing the lack of convergence of $f$ to $c$.
We can apply Proposition \ref{focussedsub} to $f$ and $w_c$, which
gives us a \proper{} subsequence $h_c:\gamma_c\to\beta_c$ such that
$g_c=f\circ w_c\circ h_c$ is a weak reduction sequence.

If $g_c$ is divergent we are done. Otherwise it converges to a limit $e$.
Again, we have a \proper{} subsequence $w_e$ witnessing that $f$ does not converge
to $e$.
From this we can construct another \proper{} weak reduction sequence $h$
that alternates between elements from $g_c$ and $f\circ w_e$. This cannot converge to $e$ because of the \proper{} overlap
with $f\circ w_e$, but if $h$ converged to any other $e'$ then because of the \proper{} overlap with
$g_c$ all neighbourhoods of $e$ and $e'$ intersect, contradicting the Hausdorff proviso.
\end{proof}
The argument of the proof is essentially the same as in \cite{kahrs-inf-modularity}, but reworded
to avoid reference to distances. In \cite{kahrs-inf-modularity} it was said that it was ``not clear''
if the proposition could be generalised to non-Hausdorff spaces.
It actually cannot, as our earlier counterexample to a generalisation
of Proposition \ref{weakconverge} would also a refute a generalisation of Proposition \ref{focussedprop}.

\section{Finite Terms and Infinitary Terms}
A \emph{signature} is a pair $\Sigma=(\mathcal{F},\#)$ where $\mathcal{F}$ is a set (of function symbols)
and $\#:\mathcal{F}\to\Nat$ is the function assigning each symbol its arity. We assume an infinite set
$\Var$ of \emph{variables}, disjoint from $\mathcal{F}$.
The set of finite terms over $\Sigma$ is called $\Ter$ and it is defined
to be the smallest set such that (i) $\Var\subset\Ter$ and (ii) $F(t_1,\ldots,t_n)\in\Ter$ whenever
$F\in\mathcal{F}\wedge\#(F)=n\wedge\{t_1,\ldots,t_n\}\subset\Ter$.
The \emph{root symbol} of a term $F(t_1,\ldots,t_n)$ is $F$, the root symbol
of a variable $x$ is $x$.

We write $\Nat^*$ for the free monoid over the natural numbers (i.e.\ finite words),
with neutral element $\emptypos$ and infix $\cdot$ as monoid multiplication.
The set of positions $\npos{t}$ of a finite term $t$ is the smallest subset of $\Nat^*$ such that:
(i) $\emptypos\in\npos{t}$, (ii) $\{ i\cdot q \mid 1\leq i\leq n, q\in\npos{t_i}\}\subset \npos{F(t_1,\ldots,t_n)}$.
Positions are used for selecting or replacing subterms.
For selecting the subterm at position $p$ of a term $t$ we use the notation $\subterm{t}{p}$.
We extend the domain of this function by setting $\subterm{t}{p}=x$ if $p\notin\npos{t}$ which allows
to extend this notation pointwise to sequences of terms: $(\subterm{f}{p})(\alpha)=\subterm{f(\alpha)}{p}$.
For replacing the subterm at position $p$ of a term $t$ by $u$ we use the notation $t[u]_p$;
if $p\notin\npos{t}$ then we set $t[u]_p=t$.
Positions are partially ordered by the \emph{prefix-ordering}: $p\preceq p\cdot q$.

A $\Sigma$-\emph{algebra} is a set $A$ together with functions $F_A:A^n\to A$ for every
$F\in\mathcal{F}$ with $\#(F)=n$. A \emph{valuation} into $A$ is a function $\rho:\Var\to A$.
Any $\Sigma$-algebra $A$ determines an interpretation function
$\sem{\_}_A:\Ter\times(\Var\to A)\to A$ as follows:
\begin{eqnarray*}
\sem{x}_A^\rho &=&\rho(x),\qquad\text{if}~x\in\Var\\
\sem{F(t_1,\ldots,t_n)}_A^\rho&=&F_A(\sem{t_1}_A^\rho,\ldots,\sem{t_n}_A^\rho)
\end{eqnarray*}
Given two $\Sigma$-algebras $A$ and $B$, a $\Sigma$-algebra \emph{homomorphism}
from $A$ to $B$ is a function $h:A\to B$ such that $h(F_A(a_1,\ldots,a_n))=
F_B(h(a_1),\ldots,h(a_n))$. If $h:A\to B$ is a homomorphism then
$h(\sem{t}_A^\rho)=\sem{t}_B^{h\circ\rho}$.

A \emph{context} is a pair of a term $t$ and a position $p\in\npos{t}$. If $C[~]$ is the context
$(t,p)$ and $u$ a term the context application $C[u]$ is defined as $t[u]_p$. A substitution
is a function $\theta:\Var\to\Ter$ which is uniquely
extended to a function $\overline{\theta}:\Ter\to\Ter$ by requiring that $\overline{\theta}$
is a $\Sigma$-algebra homomorphism.

An \emph{ultra-metric map} (short: umm) is an $n$-ary function $f:[0,1]^n\to [0,1]$ such that
(i) $f(x_1,\ldots,x_n)=0\iff x_1=0\wedge\cdots\wedge x_n=0$ and (ii) it distributes over max:
$f(\max(x_1,y_1),\ldots,\max(x_n,y_n))=\max(f(x_1,\ldots,x_n),f(y_1,\ldots,y_n))$.
Each $n$-ary umm $f$ can be expressed in the form $f(x_1\ldots,x_n)=\max_{1\leq i\leq n}\tilde{f}_i(x_i)$,
where the functions $\tilde{f}_i:[0,1]\to[0,1]$, called the \emph{components} of $f$, are also umms.
A \emph{term metric} for a signature $\Sigma$ is a $\Sigma$-algebra $m$ with carrier set $[0,1]$,
such that for each $F\in\mathcal{F}$ the function $F_m$ is a umm of the same arity.
A term metric is called \emph{continuous} \textiff{} all its ultra-metric maps are continuous.
In this paper, we shall only consider continuous term metrics.

For every term metric $m$, the distance function $d_m:\Ter\times\Ter\to [0,1]$ is defined as follows: $d_m(t,t)=0$,
$d_m(t,u)=1$ if $t$ and $u$ have different root symbols, and $d_m(F(t_1,\ldots,t_n),F(u_1,\ldots,u_n))=
F_m(d_m(t_1,u_1),\ldots,d_m(t_n,u_n))$. Some fundamental results about term metrics from
\cite{kahrs-convergence-acta} are: $d_m$ is an ultra-metric; the topology induced by $d_m$ is discrete.
Moreover, the set of \emph{infinitary terms} over $\Sigma$ and $m$ is called $\Term{m}$ and defined
as the metric completion of the metric space $(\Ter,d_m)$. $\Term{m}$ is a $\Sigma$-algebra if $m$ is continuous.

Each position $p$ in $\npos{t}$ has a unique umm $(t,p)_m$ associated with a term metric $m$:
\begin{eqnarray*}
(t,\emptypos)_m&=&\mathit{id}\\
(F(t_1,\ldots,t_n),i\cdot q)_m &=& \tilde{F}_{m,i} \circ (t_i,q)_m
\end{eqnarray*}
This associated umm gives us the distance function for position $p$,
in the sense that: $d_m(t[u]_p,t[s]_p)=(t,p)_m(d_m(u,s))$. As the definition only depends on a finite
part of $t$ it is even defined for infinite terms $t$ which are not in $\Term{m}$.

The notions of contexts, substitutions, positions and the operations on them lift canonically from
$\Ter$ to $\Term{m}$, at least for continuous term metrics, see \cite{kahrs-convergence-acta}. 

\begin{prop}\label{context-nonexp}
The subterm replacement operation $t[u]_p$ is non-expansive in $t$, for any term metric.
\end{prop}
\begin{proof}
By induction on the length of $p$.
Base case: $d_m(t[u]_\emptypos,t'[u]_\emptypos)=d_m(u,u)=0\leq d_m(t,t')$.
Otherwise $p=i\cdot q$.
If $t$ and $t'$ have different root symbols then
$d_m(t,t')=1\geq d_m(t[u]_p,t'[u]_p)$.
Finally, let $t=F(t_1,\ldots,t_n)$ and $t'=F(t_1',\ldots,t_n')$. Then
\[d_m(t,t')=\max_{1\leq j\leq n} \tilde{F}_{m,j}(d_m(t_j,t'_j))=\max(\delta,\tilde{F}_{m,i}(d_m(t_i,t'_i)))\]
where $\delta=\max_{1\leq j\leq n, j\neq i} \tilde{F}_{m,j}(d_m(t_j,t'_j))$. 
Moreover, \[d_m(t[u]_{i\cdot q},t'[u]_{i\cdot q})=\max(\delta,\tilde{F}_{m,i}(d_m(t_i[u]_q,t'_i[u]_q)))\]
Induction hypothesis on $q$ gives us:
$d_m(t_i,t'_i)\geq d_m(t_i[u]_q,t'_i[u]_q)$, so the result follows by monotonicity
of the functions $\tilde{F}_{m,i}$ (as they are umms) and $\max$.
\end{proof}
Aside: subterm replacement $t[u]_p$ is non-expansive in $u$ if and only if the term metric $m$
is non-expansive, i.e.\
if all functions $F_m$ for $F\in\mathcal{F}$ are non-expansive.

\subsection{Characterising Convergence in Term Metric $m$}

The $\epsilon$-positions of a term $t$ w.r.t.\ to term metric $m$, the set $\epos{t}$,
is defined as $\{p\in\npos{t}\mid (t,p)_m(1)\geq\epsilon\}$. Clearly, $\epos{t}$ is prefix-closed:
if $q\cdot w=p$ then $(t,q)_m(1)\geq(t,q)_m((\subterm{t}{q},w)_m(1))=(t,q\cdot w)_m(1)\geq\epsilon$.

The properties of $\epos{t}$ can be used to characterise whether an infinite term
(from $\Terinf$) exists (in $\Term{m}$).

\begin{prop}\label{finite-epos}
If $t\in\Term{m}$ then $\epos{t}$ is always finite.
\end{prop}
\begin{proof}
Let $t_\epsilon\in\Ter$ be a finite term such that $d_m(t,t_\epsilon)=\eta_1<\epsilon$ --- such finite terms
always exist, by construction of the metric completion.

Let $p\in\epos{t}$ and $y\in\Var$ be fresh. Then $d_m(t,t[y]_p)=(t,p)_m(1)\geq\epsilon$, by definition of
$\epos{t}$. We also have $d_m(t[y]_p,t_\epsilon[y]_p)=\eta_2\leq\eta_1<\epsilon$,
because subterm replacement is non-expansive.
Thus, using the ultra-metric properties of $d_m$, we know
$\epsilon\leq (t,p)_m(1)=d_m(t,t[y]_p)\leq\max(\eta_1,\max(d_m(t_\epsilon,t_\epsilon[y]_p),\eta_2))$.
As both $\eta_i$ are smaller than $\epsilon$ this is equivalent to
$d_m(t,t[y]_p)\leq d_m(t_\epsilon,t_\epsilon[y]_p)$, and the dual argument shows that these distances
must be equal.
As that value is non-zero, $p$ must be in $\npos{t_\epsilon}$.
This shows $\epos{t}\subseteq\npos{t_\epsilon}$ and as the latter is finite, so is the former.
\end{proof}
Thus, for any infinitary term under any term metric there are only finitely many subterm positions
which --- if changed --- could change the overall term by $\epsilon$ or more. This observation plays a role
when we reason about divergence.

An immediate consequence of Proposition \ref{finite-epos} is that if $\epos{t}$
is not finite (for any $\epsilon>0$) then $t\notin\Term{m}$. This gives us a criterion
to rule out certain infinite terms. In fact, the criterion is complete:
\begin{prop}\label{inf-criterion}
Let $t\in\Terinf$. If $\epos{t}$ is finite, for all $\epsilon$, then $t\in\Term{m}$.
\end{prop}
\begin{proof}
We need to create a Cauchy-sequence (w.r.t.\ $d_m$) of finite terms converging to $t$.
To make the $n$-th term $t_n$ of the sequence to be within distance of $1/n$ of the limit $t$
we obtain it by replacing all subterms of $t$ in minimal (w.r.t.\ $\preceq$) positions not in $\epos{t}$
(for $\epsilon=1/n$) by the variable $x$. As $\epos{t}$ is finite its elements have a maximal length $k$
and the longest position in $t_n$ is (at most) of length $k+1$,
and thus by K\"onig's Lemma $t_n$ must be finite too.
\end{proof}

An important concept when studying convergence is the equivalence relation that two
terms have ``the same function symbols up to position $p$''.
Formally, this is defined by induction on the length of the positions as follows:
\[
\begin{array}{l}
\topequ{t}{\emptypos}{u}\\
\topequ{F(t_1,\ldots,t_n)}{i\cdot q}{F(u_1,\ldots,u_n)} \Longleftarrow \topequ{t_i}{q}{u_i}
\end{array}
\]
Clearly, if $\topequ{a}{p}{b}$ then $(a,p)_m=(b,p)_m$; see also \cite{kahrs-convergence-acta}.


Traditionally, infinitary term rewriting has focussed on one particular distance function $d_\infty$.
The same distance function arises through the definition of a term metric:
we define the term metric $\infty$ in which all $n$-ary ultra-metric maps
are given as $\infty_n(x_1,\ldots,x_n)=\frac{1}{2}\cdot\max(x_1,\ldots,x_n)$.
It plays a fundamental role as $\Terinf$ includes \emph{all} infinite terms:
\begin{prop}\label{cofinal}
Any Cauchy-sequence in $(\Ter,d_m)$ is also a Cauchy-sequence in $(\Ter,d_\infty)$.
\end{prop}
\begin{proof}
Let $f:\omega\to\Ter$ be that sequence.

Assume it was not Cauchy under metric $d_\infty$. Then there is a maximal (under the prefix ordering)
position $p$ such that $\subterm{f}{p}$ changes its root symbol infinitely often.
By maximality, the symbols above $p$ are fixed for all sufficiently large $k$,
i.e.\ $\topequ{f(k)}{p}{f(k+1)}$, and infinitely often have $\subterm{f(k)}{p}$ and
$\subterm{f(k+1)}{p}$ different root symbols.

This implies (see proposition 6 in \cite{kahrs-convergence-acta}) that there is a fixed
umm $C_m$ such that for all sufficiently large $k$
\[d_m(f(k),f(k+1))\geq C_m(d_m(\subterm{f(k)}{p},\subterm{f(k+1)}{p})).\]
Whenever we have a change of function symbol at position $p$ the expression on the right becomes the fixed value
$C_m(1)$. As this happens infinitely often we have a contradiction to the
pre-condition that $f$ is Cauchy under metric $d_m$.
Therefore the assumption was wrong and $f$ is Cauchy under $d_\infty$ as well.
\end{proof}
\begin{cor}\label{idextend}
For any term metric $m$, there is a unique continuous extension of the identity function from finite terms
to $\Term{m}\to\Terinf$.
\end{cor}
\begin{proof}
The identity is continuous on finite terms between any two term metrics, because the
topologies on finite terms are discrete for any term metric (proposition 4 in \cite{kahrs-convergence-acta}).
Proposition \ref{cofinal} allows to extend the function to infinite terms with domain $\Term{m}$,
and such extensions to metric completions are always unique when they exist.
\end{proof}
Proposition \ref{cofinal} shows that $\Terinf$ ``contains all infinite terms'', because
any infinite term arises as the limit of a Cauchy-sequence under some metric $d_m$,
and the corollary indicates that the construction is continuous.
Thus we can view any element of $\Term{m}$ also as an element of $\Terinf$.
It also means that any sequence in $\Term{m}$ that fails to converge under $d_\infty$ will
necessarily also fail to converge under $d_m$.

Aside: for finite signatures there is a simpler argument to establish Corollary \ref{idextend}:
the identity function is uniformly continuous (from $d_m$ to $d_\infty$). However,
this is no longer true for infinite signatures, e.g.\ if a signature contained
infinitely many unary function symbols $F\langle n\rangle$ with umms $F\langle n\rangle_m(x)=\frac{x}{n}$.

\begin{prop}\label{limitexists}
Let $f$ be a sequence over $\Term{m}$ such that $f$ is converging under metric $d_\infty$
to $t\in\Term{m}$. Then $f$ is also converging to $t$ under metric $d_m$.
\end{prop}
\begin{proof}
As $t\in\Term{m}$, the set $\epos{t}$ is finite for any $\epsilon$.
As $f$ is converging to $t$ under $d_\infty$, there must be a $\gamma_0$ such
that $\forall\gamma\geq\gamma_0.\:\forall p\in\epos{t}.\:\topequ{t}{p}{f(\gamma)}$.
This implies that $t$ and all $f(\gamma)$ are within $\epsilon$-distance, w.r.t.\ to metric $d_m$.
Thus $f$ stays eventually within the $\epsilon$-ball of $t$, and $f$ must converge to $t$ as this holds for any
$\epsilon$.
\end{proof}
Therefore, if a sequence of finite terms $f$ converges under metric $d_\infty$ to a limit
$t\in\Terinf$, the question whether $f$ converges under $d_m$ reduces to the problem whether
$t\in\Term{m}$ or not.
If $t\in\Term{m}$, Proposition \ref{limitexists} applies and so $f$ converges; if $t\notin\Term{m}$
then $f$ cannot converge under $d_m$, because if it did converge to $t'\neq t$ then
$f$ would converge under $d_\infty$ to both $t$ and $t'$ --- which is impossible in a Hausdorff space.

Let $t\in\Terinf$. An \emph{infinite path} of $t$ is a strictly monotonic function $f:\omega\to\npos{t}$, i.e.\
$n<k\Rightarrow f(n)\prec f(k)$. The path $f$ is called \emph{full} \textiff{} the length of $f(n)$ is always $n$.
We say that $f$ converges w.r.t.\ term metric $m$ \textiff{} the sequence of umms
$(t,f(n))_m$ converges to the constant 0 function.

\begin{lem}\label{path-lemma}
Let $f$ be an infinite path of $t$. Then $f$ converges w.r.t.\ $m$ \textiff{}
the sequence $(t,f(n))_m(1)$ converges to $0$.
\end{lem}
\begin{proof}
If $(t,f(n))_m$ converges to the constant 0 function then it will converge to 0 when applied to argument 1.
If $(t,f(n))_m(1)$ converges to $0$ then by monotonicity $(t,f(n))_m(\epsilon)$ must converge to $0$ at
(at least) the same rate, and therefore $(t,f(n))_m$ converges to the constant 0 function.
\end{proof}

\begin{prop}\label{infinite-paths}
Let $t\in\Terinf$. Then $t\in\Term{m}$ \textiff{} all infinite paths of $t$ converge w.r.t.\ term metric $m$.
\end{prop}
\begin{proof}
Suppose path $f$ does not converge. Then by \ref{path-lemma} the sequence
$(t,f(n))_m(1)$ does not converge to $0$. Thus, $(t,f(n))_m(1)\geq\epsilon>0$ for infinitely many $n$.
As each of these is in $\epos{t}$, this set is infinite and $t\notin\Term{m}$ follows
from Proposition \ref{finite-epos}.

Now assume $t\notin\Term{m}$. Then by proposition \ref{inf-criterion} $\epos{t}$ must be infinite
for some $\epsilon$. We can iteratively construct a full non-converging infinite path through
$\epos{t}$ by the pigeon hole principle:
set $f(0)=\emptypos$; as the symbol at $\subterm{t}{f(n)}$ has finite arity,
there must be at least one $i$ such that
infinitely many positions in $\epos{t}$ have $f(n)\cdot i$ as prefix, and we can set
$f(n+1)=f(n)\cdot i$. As all values in the range of $f$ are in $\epos{t}$,
the sequence $(t,f(n))_m(1)$ stays above $\epsilon$ and is therefore not converging to 0.
\end{proof}

\section{Infinitary Term Rewriting}
An \emph{infinitary term rewriting system} (iTRS) is a triple $(\Sigma,m,R)$ such that
(i) $\Sigma$ is a signature, (ii) $m$ is a term metric for $\Sigma$, and (iii)
$R\subseteq\Term{m}\times\Term{m}$ is a set of rewrite rules such that
for all $(l,r)\in R$ we have $l\notin\mathit{Var}$ and all variables in $r$ also occur in $l$.
We write $l\to r$ if $(l,r)\in R$.

Note: this definition permits infinite terms on both left-hand and right-hand sides of rules,
as well as repeated variables in the left-hand side.
For better continuity properties
of the rewrite relation one would typically require in addition that each $l$ is a finite linear term.
Moreover, \cite{kahrs-convergence-acta} required
all rules to be depth-preserving (see below).
However, for the purposes of this paper these restrictions are not necessary.

A relation $S$ on $\Term{m}$ is called \emph{substitutive} \textiff{} $s\mathrel{S}t$ implies 
$\overline{\theta}(s)\mathrel{S}\overline{\theta}(t)$,
for any substitution $\theta$.
It is called \emph{compatible} \textiff{} $s\mathrel{S}t$ implies $C[s]\mathrel{S}C[t]$ for any context $C[~]$.

The single-step rewrite relation $\to_R$ of an iTRS $(\Sigma,m,R)$ is
defined as the compatible and substitutive closure of $R$, i.e.\
$C[\overline{\theta}(l)]\to_R C[\overline{\theta}(r)]$ whenever $(l,r)\in R$,
$C[~]$ is a context, and $\theta$ is a substitution.

Aside: there is a slight issue with the definition
of ``compatible''. For better continuity properties one
could require in addition that a compatible relation should be reflexive on infinite terms.
If $t$ is an infinite term then a rewrite step
$t[l]_p\to_R t[r]_p$ should vanish in the limit, as $p$ goes arbitrarily deep.
This is necessary to ensure that the relation $\to_R$ is upper semi-continuous, and that it is closed.
For the purposes of this paper the distinction is immaterial as the presence of reflexive rewrite steps
does not affect the convergence relations.

We associate with an iTRS $(\Sigma,m,R)$ the MARS $(\Term{m},d_m,\to_R)$; in particular, the iTRS
is called convergent \textiff{} its associated MARS is.

In the following we will be using a generalisation of the notion of
\emph{depth-preservation} \cite{inf-orth} to arbitrary term metrics.
In term metric $\infty$, the depth of a variable $x$ in a term $t$ is the length $l$ of the shortest
position $p$ in $t$ such that $\subterm{t}{p}=x$. Equivalently, that information is also present
in the distance $d_{\infty}(t,t[y/x])=(t,p)_\infty(1)$ (for fresh $y$), because that distance is $2^{-l}$.

For other term metrics that figure is insufficient to fully capture the notion of depth of a variable
in a term, because it only tells us what happens with distances when terms with different roots
replace the $x$. More generally, we are interested in the distances $d_m(t[u/x],t[s/x])$ for varying $u$ and $s$;
the dependence on $u$ and $s$ can be replaced by a dependence on $d_m(u,s)$.
Thus the generalised \emph{variable depth} of $x$ in $t$ w.r.t.\ term metric $m$,
$\mathit{vdepth}(x,m,t)$, is an ultra-metric map of type $m\to m$, given as:
\[\mathit{vdepth}(x,m,t)(y)=\sem{t}^{\rho(y)}_{m}\quad\text{where}~\rho(y)(z)=
\begin{cases}
y & \text{if}~z=x\\
0 & \text{otherwise}
\end{cases}
\]
A rewrite step $t\to u$ preserves the depth of $x$ \textiff{}
$\mathit{vdepth}(x,m,t)\geq\mathit{vdepth}(x,m,u)$, in the pointwise order on $\Var\to m$.
Overall, a relation $R$ on $\Term{m}$ is called
\emph{depth-preserving} \textiff{} $\forall t,u.\:t\mathrel{R}u\Rightarrow\sem{t}\geq\sem{u}$ which means that
all its rewrite steps preserve the depth of all variables.
Here, the order $\geq$ is the pointwise order on the function space, inherited (twice) from $m$.

An iTRS (for term metric $\infty$) is \emph{strongly converging} \textiff{} for all open reduction sequences of length $\alpha$
no redex position $p$ is reduced arbitrarily close to $\alpha$. Equivalent to this \cite{zantema-inf-normalisation}
is the notion of top-termination, i.e.\ that no reduction sequence contracts a root redex infinitely often.

It is possible to define the property without referring to redex positions though:
the \emph{indirected version} of an iTRSs $(\Sigma,m,R)$ is defined as follows.
Its signature is $\Sigma$ extended by a fresh unary
function symbol $I$; for each rule $l\to r$ in $R$ it has a rule $l\to I(r)$; in addition,
it has the rule $I(x)\to x$.  We extend the term metric $m$ by setting $I_m(x)=x$.


\begin{prop}
An iTRS over term metric $\infty$ is \emph{strongly converging} \textiff{} its indirected version is converging.
\end{prop}
\begin{proof}
Any reduction sequence $f$ of the original iTRS can be mapped into a reduction sequence
$f'$ of the indirected version by splitting the step into two parts, where the second part removes
the just introduced $I$-symbol. If $f$ is not top-terminating then $f'$ fails to converge,
as it would have infinitely many changes of root-symbol.

Dually, a reduction sequence $g$ of the indirected version can be mapped into a (shorter) sequence
$g^{+}$ of the original system, by removing all $I$-symbols. However, $g$ can only fail to converge
by an infinitely often occurring function symbol change at some minimal position $p$.
That means though that $g$ must apply rules other than $I(x)\to x$ at position $p$ infinitely often,
and thus $g^{+}$ would apply those rules at $p$ or higher infinitely often which means that $g^{+}$
is not strongly converging.
\end{proof}

The reason that this proposition is limited to term metric $\infty$ is simply that the notion of strong convergence
is only \emph{defined} for metric $d_\infty$. The idea behind strong convergence is that redex contractions
must contribute less and less to the diameter of a sequence as redex positions go deeper.
For other term metrics the movement of the redex positions alone does not suffice,
similarly as modifications of the metric for the infinitary $\lambda$-calculus \cite[page 709f]{terese-inf}
affect the notion of ``root-active''.

For other term metrics than $\infty$ \emph{we define} strong convergence as the convergence of the indirected iTRS.

Consequence: consider the term metric $\mathit{id}$ with $F_\mathit{id}(x_1,\ldots,x_n)=\max(x_1,\ldots,x_n)$
for all function symbols $F$.
This is a discrete metric
and there are no infinite terms: $\Term{\mathit{id}}=\Ter$. Thus, convergence means that reduction sequences
are eventually constant, and strong convergence coincides with strong normalisation.

This also shows that strong convergence is not in general a modular property,
simply because strong normalisation is not. However, strong normalisation is modular
in the absence of collapsing rules \cite{modular-term}, so we will consider
the corresponding constraint here.

\section{Examples, Counterexamples and Near Counterexamples}
In this section we will be looking at examples that show
why the modularity results for convergence
come with side conditions. The examples here are using metric $d_\infty$; to consider
other metrics at all we would have to first explain what the disjoint union of two iTRSs
with arbitrary metrics is,
see the next section.

\subsection{Collapsing Rules}
Particularly problematic w.r.t.\ convergence are collapsing rules, i.e.\ rules that
have a variable as their right-hand side. An iTRS immediately fails to be strongly convergent
if a collapsing rule is present. The reason is this: any such rule takes the form
$C[x,\ldots,x]\to x$, the equation $t=C[t,\ldots,t]$ has a (unique) solution in $\Terinf$, and
$t\stackrel{\emptypos}{\to} t$ gives an infinite reduction sequence with contractions at the root.

The weaker notion of convergence can coexist with collapsing rules, but barely:
if $C[x]\to^{*}x$ holds for any other contexts $C[~]$ than those of the form $C[y]=F^n(y)$
(for varying $n$ but fixed $F$) then such an iTRS is not convergent either.
In particular, the presence of two rules $F(x)\to x$ and $G(x)\to x$ means that the terms $t=F(G(t)), u=G(F(u))$ reduce
to each other and thus give rise to a non-converging reduction sequence.
The example also shows that convergence (on its own) is not a modular property.

Convergence can also be broken if only one of the two systems has a collapsing rule.
\begin{exa}
System $R$ has rule $G(H(x))\to G(x)$, system $S$ has the rule $F(x)\to x$.
In the combined system there is the term $t=F(H(t))$ and the non-convergent reduction
$G(t)\to G(H(t))\to G(t)$.
\end{exa}

\subsection{Building Infinite Results}\label{bir}
Typically we can observe convergence (or its absence) at $\omega$-indexed reduction sequences.
In the presence of non-left-linear rules this can be rather different though. 
\begin{exa}\label{ex-zantema} (from \cite{zantema-inf-normalisation})
$\{ {E\to S(E)},~ {F\to S(F)},~ {G(x,x)\to G(E,F)} \} $

\noindent This fails to converge, but this failure first occurs at $\omega^2$-indexed reduction sequences.
We have $G(E,F)\to^{*} G(S^n(E),S^m(F))$ for any $m,n$ and in the limit
$G(E,F)\weak G(S^\infty,S^\infty)\to G(E,F)$. This does not provide any convergence problems
at ordinals of the form $\omega\cdot k$ though; however, when we attempt to extend a reduction sequence
to $\omega^2$ we find both $G(E,F)$ and $G(S^\infty,S^\infty)$ in the image of the reduction sequence
at any of its neighbourhoods.
\end{exa}

This is a common pattern for establishing non-convergence: any sequence of length $\alpha$
that begins and ends in the same value, but that visits a different value in between,
gives rise to a non-converging sequence of length $\omega\cdot\alpha$ by simply repeating
that sequence $\omega$ times. That construction works in any Hausdorff space.

\begin{exa}\label{rearrange}
In Example \ref{ex-zantema} we had $E\to^*S^n(E)$ for any $n$, and $E\weak S^\infty$ in the limit.
However, one can have an iTRS with the former but not the latter property,
e.g.\ if we replace the first rule by these four:
\[\{ {E\to Z},~{E\to H(E)},~{H(Z)\to S(Z)},~H(S(x))\to S(S(x))\}\]
Patterns like this complicate the reasoning about (weak) convergence, because in a combined system
an outer layer may place in a position $p$ different reducts of $E$, and thus the term in position
$p$ could converge at a limit to $S^\infty$, showing that a principal subterm at a limit may not be a reduct
of an earlier principal subterm. This happens when we combine this system with the following:
\begin{align*}
\{ J(K(x,y)) & \to J(y) \}
\end{align*}
Let $t=K(E,t)$ and $u=K(S^\infty,u)$: we have
$t=K(E,K(E,\ldots))\weak K(Z,K(S(Z),\ldots))$, and thus
$J(t)\weak J(K(Z,K(S(Z),\ldots)))$. Continuing the reduction from there
by repeatedly applying the rule $J(K(x,y)) \to J(y)$ we have overall $J(t)\weak J(u)$.

What is remarkable about this is that
$E$ is the only principal subterm of $J(t)$ and it does not reduce to $S^\infty$,
the only principal subterm of $J(u)$.
However, this second iTRS already fails to converge: given the term $s=K(J(K(x,y)),s)$ we can construct the
non-converging sequence $J(s)\to J(K(J(y)),s)\to J(s)$.
\end{exa}

This construction is by no means arbitrary, as we shall see later:
if a reduction sequence $f$ in the combined system ``substantially rearranges'' principal subterms
then one can construct a non-converging reduction sequence in the iTRS to which the root symbol belongs,
where the role of principal subterms is played by some fixed terms $l\neq r$ with $l\to r$.

\subsection{Beyond Ordinary iTRSs}
What happens to convergence if we remove any constraints on rewrite rules, i.e.\ if we allow
left-hand sides to be variables (expansive rules),
and right-hand sides to contain extra variables the left-hand sides does not?

One of the main reasons these kind of rules are excluded in finite term rewriting 
is that either kind of rule immediately breaks the strong normalisation property.
Another reason is more pragmatic: the presence of such rules requires
a very different implementation.

Expansive rules immediately break strong convergence, but they may not break weak convergence.
The rule $x\to F(x)$ on its own is weakly convergent,
and if disjointly added to a strongly convergent iTRS
the result would be weakly convergent. On the other hand, the presence of a second expansive
rule $y\to G(y)$ breaks weak convergence: $x\to F(x)\to G(F(x))\to F(G(F(x)))\to\ldots$.
Similar to collapsing rules, the presence of two different function symbols in an expansive rule breaks convergence:
in that case the right-hand side must have the form $C[F(t_1,\ldots,t_n)]_p$
where the root symbol of $C[~]$ is different from $F$.
The reduction sequence that applies the rule alternatively
in positions $\emptypos$ and $p$ is non-converging, as the function
symbol in position $p$ changes every time.

Rules with extra variables may not only be convergent, they may even be strongly convergent.
An example for that is the rule $F(x)\to G(y)$. This is not only convergent on its own, it preserves
convergence if disjointly added to any convergent iTRS. A rule like $l\to y$, where the right-hand side \emph{is}
an extra variable, can only be convergent if $l$ is a ground term, as otherwise two different
substitution instances of $l$ could rewrite to each other. If $l$ is ground then $l\to y$ can only be convergent
for (metrics homeomorphic to) $d_\infty$, as the rule can be used to gradually build terms of any infinite shape.

But that does not mean that extra variables are unproblematic:
rules with extra variables always fail to be depth-preserving, for any term metric.
They also break many proof techniques, as they enable the rank of terms to go up by rewriting.






\section{The Disjoint Union of iTRSs}
A \emph{signature morphism} between signatures $(\mathcal{F}_1,\#_1)$ and $(\mathcal{F}_2,\#_2)$
is a function $\sigma:\mathcal{F}_1\to\mathcal{F}_2$ that preserves arities.
This gives rise to a translation between finite terms $\overline{\sigma}$ defined
as $\overline{\sigma}(F(t_1,\ldots,t_n))=\sigma(F)(\overline{\sigma}(t_1),\ldots,\overline{\sigma}(t_n))$, and
$\overline{\sigma}(x)=x$ for $x\in\Var$.

A \emph{metric signature morphism} between $(\Sigma_1,m_1)$ and $(\Sigma_2,m_2)$
is a signature morphism $\sigma:\Sigma_1\to\Sigma_2$ such that for all $F\in\mathcal{F}_1$
we have: $F_{m_1}=\sigma(F)_{m_2}$.
For a metric signature morphism $\sigma$ the function $\overline{\sigma}$ is a homeomorphic embedding
as it preserves distances: $d_{m_2}(\overline{\sigma}(t),\overline{\sigma}(u))=d_{m_1}(t,u)$,
and therefore it extends to infinitary terms.

Notice that signature morphisms preserve the \emph{term metric} $m$, not just the derived notion of the metric $d_m$.
Different term metrics can give rise to identical topologies:
\begin{exa}
If $\Sigma$ comprises a single unary function symbol $F$ then both term metrics
$F_\mathit{id}(x)=x$ and $F_r(x)=\sqrt{x}$
give rise to identical metrics and therefore produce the
same set $\Term{\mathit{id}}=\Term{r}=\Ter$, with no infinite terms at all. 
But, combined with another unary symbol $G$
with term metric $G_\infty(x)=x/2$ the two versions of $F$ behave differently: the term
$t=G(F(t))$ exists with $F_\mathit{id}$ and does not with $F_r$.
\end{exa}

An \emph{iTRS morphism} from $(\Sigma_1,m_1,R_1)$ to $(\Sigma_2,m_2,R_2)$ is a metric signature morphism
$\sigma:(\Sigma_1,m_1)\to(\Sigma_2,m_2)$ such that $\overline{\sigma}(R_1)\subseteq R_2$, where
$\overline{\sigma}$ is canonically extended to rules and sets of rules.

Infinitary term rewriting systems and their morphisms form a cocomplete category.
The \emph{disjoint union} $R+S$ of two iTRSs $R$ and $S$ is defined to be the coproduct in that category.
This contains all symbols from both systems
(possibly renamed if there was a conflict) and all rules from both systems
whilst preserving the umms for all function symbols.
As distances are preserved by iTRS morphisms we shall use the metric $d_m$
of the coproduct also for distances of the constituent iTRSs.

The preservation of the metric by the injection into the coproduct
also settles the ``reflected by'' part of modularity properties:
\begin{thm}
The properties of convergence and strong convergence are reflected by the disjoint union of iTRSs.
\end{thm}
\begin{proof}
Convergence: assume $R+S$ is converging, let $\iota_1$ be the coproduct injection from $R$ to $R+S$.
Any open reduction sequence $f$ of $R$ is mapped by $\iota_1$ into an open reduction sequence of $R+S$
which by assumption is converging. As $\iota_1$ is an isometric embedding (and as its domain is complete)
it reflects the convergence back to $f$.

Strong convergence: any counterexample to strong convergence is clearly preserved by $\iota_1$,
as the map does not affect redex positions.
\end{proof}

A \emph{signature extension} of an iTRS $R$ is the disjoint union of $R$
with another iTRS with empty set of rules.

One might expect that signature extensions preserve most properties of interest,
but that is questionable. We can even have that a signature extension turns a strongly converging iTRS 
into one that is not converging at all.

\begin{exa}\label{exa-layers}
Consider the iTRS with rule $F(F(x))\to G(x)$ under term metric $\infty$ --- this is strongly
converging. If we extend the signature with a function symbol $H$ with associated umm $H_m(x)=\min(1,2\cdot x)$
then convergence is broken: the term $t=F(F(H(t)))$ exists under the combined metric, but repeatedly
applying the rule to this term could only give us $u=G(H(u))$ which does not exist in that metric.
\end{exa}

There are two ``culprits'' contributing to this: (i) the signature extension used an expansive umm,
(ii) the iTRS was not depth-preserving. 
In this case, the counterexample can be reconstructed with non-expansive metrics:
\begin{exa}\label{exa-layers2}
Modify Example \ref{exa-layers} by setting $F_m(x)=x^2$, $G_m(x)=x$, $H_m(x)=\min(x,\frac{1}{2})$.
In this case, neither of the two iTRSs permits
infinite terms, but their combination does; the counterexample is preserved, as
$t$ does and $u$ does not exist under this metric.
\end{exa}

\section{Auxiliary definitions and observations}
When studying the disjoint union of term rewriting systems $R$ and $S$ it is useful to split
a term into \emph{layers} of the participating systems, and distinguish its \emph{principal subterms}. 
The principal subterm positions of a term $t$, $\ppos{t}\subset\npos{t}$ have the following properties:
$p$ is principal if (i) the root of $\subterm{t}{p}$ is a function symbol belonging to a different signature
than the root of $t$; (ii) no proper prefix of $p$ is principal.
The top positions of a term $t$, $\tpos{t}\subset\npos{t}$ are defined as follows:
$\tpos{t}=\{ p\in\npos{t}\mid\forall q.\:q\prec p\Rightarrow q\notin\ppos{t}\}$. Thus top positions are either
parallel to principal subterms or they are their prefixes.

A \emph{principal path} of $t$ is an infinite path $f$ where $f(0)=\emptypos$,
$f(n+1)=f(n)\cdot p_n$ such that $p_n\in\ppos{\subterm{t}{f(n)}}$.
The modularity of convergence is tightly connected to the question whether principal paths that arise
through rewriting are convergent.

The top layer of a term is comprised of all the positions reachable from the root without a change
of signature, and is thus ``cut off'' at its principal subterm positions. This leaves the layer with gaps at
the cut-off points; therefore, the top layer is technically defined as the function that recreates a term
if suitable fill material for the gaps is provided:
The \emph{top-layer} of a term $t$ is a
function $\toplayer{t}$ with type $(\ppos{t}\to\Term{m})\to\Term{m}$ where we
write $\toplayer{t}_\xi$ for the application of $\toplayer{t}$ to a function $\xi:\ppos{t}\to\Term{m}$.
The top layer has the following properties: above principal subterm positions it agrees with $t$:
\[\forall\xi.\:\forall p\in\ppos{t}.\:\topequ{t}{p}{\toplayer{t}_\xi}\]
In positions parallel to principal subterms, $t$ and $\toplayer{t}$ coincide:
\[\forall\xi.\:\forall q\in\npos{t}.\:(\forall p\in\ppos{t}.\:q||p)\Rightarrow \subterm{t}{q}=\subterm{\toplayer{t}_\xi}{q}\]
And in principal subterm positions themselves the gaps are filled with the function $\xi$:
\[\forall p\in\ppos{t}.\:\toplayer{t}_\xi/p=\xi(p).\]
The standard notation \cite{terese-inf} for layers in finite terms is $\aliencontext{C}{t_1,\ldots,t_n}$.
Elsewhere \cite{simonsen-modularity} this has been adapted for infinitary terms as $\aliencontext{C}{t_i}_{i\in I}$,
to use some (possibly infinite) set $I$ as indexing set for those positions.
The main reason for departing from that is that the $\toplayer{t}$ notation avoids the need of a separate indexing set
by using $\ppos{t}$ itself, as $\toplayer{t}_\xi=\aliencontext{t}{\xi(p)}_{p\in\ppos{t}}$.

We can define a distance between top-layers:
$d(\toplayer{t},\toplayer{u})=d_{m}(\toplayer{t}_{kx},\toplayer{u}_{kx})$ where
$kx$ is the function that constantly returns $x$, where $x$ is a fresh variable.

\begin{lem}\label{big-context}
Let $t\in\Term{m}$. Let $f:\alpha\to\ppos{t}\to\Term{m}$ be a function such that for each $p\in\ppos{t}$
the functions $g_p:\alpha\to\Term{m}$ defined as $g_p(\beta)=f(\beta)(p)$ are converging sequences.
Then $h:\alpha\to\Term{m}$ defined as $h(\beta)=\toplayer{t}_{f(\beta)}$ is a converging sequence.
\end{lem}
\begin{proof}
Suppose $t$ is a finite term. Then $\ppos{t}$ is finite, so $f$ is essentially an $n$-tuple of converging sequences which
is also a converging sequence of $n$-tuples. Moreover, $\toplayer{t}$ is uniformly continuous, as all term-forming umms are
(Proposition 5 in \cite{kahrs-convergence-acta}), and as uniformly continuous functions preserve Cauchy-sequences the result follows.

Otherwise (i.e.\ $t$ is infinite),
for any $\epsilon>0$ there is a finite term $t_\epsilon$ such that $d_m(t,t_\epsilon)<\epsilon$.
W.l.o.g.\ we can assume $\ppos{t_\epsilon}\subseteq\ppos{t}$. Thus the previous argument applies to
$t_\epsilon$ and $f$, i.e.\ the function $h_\epsilon$ defined as $h_\epsilon(\beta)=\toplayer{t_\epsilon}_{f(\beta)}$
is a converging sequence. Moreover, as subterm replacement is non-expansive (Proposition \ref{context-nonexp})
the functions $h$ and $h_\epsilon$ are pointwise within $\epsilon$-distance. Thus $h$ stays eventually
within an $\epsilon$-ball, and --- as this holds for any $\epsilon$ --- $h$ must converge.
\end{proof}
From this we get a compositionality property for converging sequences:
\begin{cor}\label{divide-and-conquer}
Let $f:\alpha\to\Term{m}$ be a sequence such that $\toplayer{f}$ converges to $\toplayer{u}$
and $\subterm{f}{p}$ converges for any $p\in\ppos{u}$. Then $f$ converges.
\end{cor}
\begin{proof}
We can apply Lemma \ref{big-context} to $u$ and $\subterm{f}{p}$, giving a converging sequence $h:\alpha\to\Term{m}$
with $h(\beta)=\toplayer{u}_{p\mapsto \subterm{f(\beta)}{p}}$. Moreover, the distances between $h$ and $f$ are pointwise
the same as the distances between their top-layers, which means that $f$ must converge to the limit of $h$.
\end{proof}

The (possibly infinite) rank of an infinitary term $t$ is defined as follows:
\[ \mathit{rank}(t) = \sup\{ 1+ \mathit{rank}(\subterm{t}{p})\mid p\in\ppos{t}\}\]


From now the reasoning will be about the disjoint union of two non-collapsing
iTRSs $R$ and $S$ with rewrite relation $\to_{RS}$ and combined signature $\Sigma$,
and for the top layer of any reduction (step or sequence) it is assumed that it is situated in system $R$,
with signature $\Sigma_R$.

In the following we will be using a construction in which all principal subterms of a term $t$ at finite
rank $n$ are replaced by a fixed term $u$, $t[n\searrow u]$. Formally:
\begin{align*}
t[0\searrow u] &= u\\
t[n+1\searrow u] & = \toplayer{t}_\xi\quad\mathrm{where}~\xi(p)=(\subterm{t}{p})[n\searrow u]
\end{align*}
This notation is also extended to sequences of terms: if $f:\alpha\to\Term{m}$ is continuous then
$f[n\searrow u]:\alpha\to\Term{m}$ is defined pointwise as $f[n\searrow u](\beta)=f(\beta)[n\searrow u]$.
Notice that the function $t\mapsto t[n\searrow u]$ is non-expansive
and therefore preserves the continuity of the sequence.

\begin{lem}\label{cutoff}
Let $f:\alpha\to\Term{m}$ be a reduction sequence. Let $u\in\Term{m}$ and $n\in\mathcal{N}$
be arbitrary. Then the sequence $f[n\searrow u]$ is a reduction sequence of the reflexive closure of $\to_{RS}$.
\end{lem}
\begin{proof}
The absence of collapsing rules means that principal subterm replacement at level $n$ commutes
with reduction steps at level $m<n$, even for contraction of non-left-linear redexes
(even if rules with extra variables were permitted).
Limits are also preserved because the operation is non-expansive.
Reduction steps at lower level become reflexive steps.
\end{proof}
Any reduction sequence of the MARS $(M,d,\to^{=})$ can be turned into a (possibly shorter)
reduction sequence of $(M,d,\to)$. Thus, the presence of reflexive steps in $f[n\searrow u]$ is merely a technicality.
The full generality of Lemma \ref{cutoff} will be used later;
but of special interest is the case $n=1$:
\begin{cor}\label{rep1}
Let $f:\alpha\to\Term{m}$ be a reduction sequence. Let $u\in\Terminfr{m}$.
Then $f[1\searrow u]$ is a reduction sequence of the reflexive closure of $\to_R$.
\end{cor}
\begin{proof}
As all terms of the sequence are by construction $\Sigma_R$-terms each $\to_{RS}$-step is a $\to_R$-step.
\end{proof}
We define the principal positions of a reduction sequence $f:\alpha\to\Term{m}$ as follows:
\[\ppos{f}=\bigcup_{\beta<\alpha}\bigcap_{\beta<\gamma<\alpha}\ppos{f(\gamma)}\]

\begin{lem}\label{toplayer}
Let $f:\alpha\to\Term{m}$ be an open reduction sequence. If $R$ is convergent
then the top layers of $f$ converge to some top layer $\toplayer{u}$ where
$\ppos{u}=\ppos{f}$.
\end{lem}
\begin{proof}
Since the iTRS of the top layer is convergent then so is $f[1\searrow x]$,
for some fresh variable $x$, and let its limit be $t$.
Let $c$ be any non-variable term in the other iTRS then by substitutivity
of weak convergence we also have $f[1\searrow x][c/x]=f[1\searrow c]\weak t[x/c]$,
and we can set $u=t[x/c]$ since $\toplayer{u}$ does not depend on the choice of $c$.
Clearly, any principal position in $t[x/c]$ must be principal for $f(\beta)[1\searrow c]$
for all $\gamma$ in the open interval $(\beta,\alpha)$ for some $\beta$.
\end{proof}

\section{Modularity of convergence}
In Example \ref{rearrange} we have
seen how a top-layer can through rearrangement of its principal subterms interfere
with the convergence at subterm positions.
However, the example that rearranged principal subterms ``in a substantial way'' was divergent.

We will show that this is not an accident.
The idea behind this is the following: if $f$ is an open reduction sequence then so is $g=f[1\searrow l]$
and if $l$ is a term belonging to the (converging) top-layer then $g$ must converge.
Moreover, if $l\to r$ (in the interesting case $l\neq r$) then $g$ must still converge if we interleave
it with the occasional reduction of $l\to r$, provided we preserve non-left-linear redexes along the way.

It is impossible to have both $l$ and $r$ occur infinitely often in a fixed position $p$,
because that would create a divergence with diameter of at least $d_\infty(l,r)\cdot 2^{-|p|}$ in
$\Terinf$, which by Proposition \ref{cofinal} gives divergence for $\Term{m}$ as well.
That means that two situations are ruled out: (i) that the top-layer reduction drags eventually infinitely
many different subterms into position $p$ --- as in $f[1\searrow l]$ each $l$ could be reduced
once it appears in position $p$; (ii) at the same time infinitely many terms appearing in position $p$
are ``descendants'' of a principal subterm $t$ and infinitely many others are not. In that case we would create
a non-converging sequence by replacing all weak reducts of $t$ in $f$ by $r$, and all other principal subterms by $l$.
With both these scenarios ruled out one can show that the sequence of subterms in position $p$ must be a focussed
sequence.

These ideas will be formalised in a slightly different way
that avoids the tracing of redexes and the rather cumbersome notion of ``descendant'' ---
which is impossible anyway for non-strong reductions \cite{simonsen-residual}.

\subsection{Predicate Sequences}
A \emph{predicate sequence} is an ordinal-indexed sequence of predicates $s$ on terms
such that ${{\beta\leq\gamma}\wedge t\weak u\wedge s(\gamma)(u)}\Rightarrow {s(\beta)(t)}$. 
A predicate sequence is meant to be used alongside a reduction sequence of the same length,
w.r.t.\ the same signature and term metric.
The idea behind predicate sequences is that they are used as a tool to decide
when to reduce an $l$ to $r$ in $f[1\searrow l]$; or rather: some principal subterms will
be replaced by $l$, some by $r$, and the construction makes sure that reduction steps are preserved.
So the motivation behind the definition of a predicate sequence is this:
once we have decided to replace a principal subterm $t$ by $r$ (rather than $l$)
then (i) we cannot change our mind \emph{later}, i.e.\ for a
larger indexing ordinal, and (ii) all ``descendants'' of $t$ need to be replaced by $r$ too, which we can
safely over-approximate as all \emph{weak reducts} of $t$.

Let $f:\alpha\to\Term{m}$ be a reduction sequence and $p$ be an $\omega$-sequence of positive integers.
The function $f_p:\alpha\to\Term{m}\to\mathcal{B}$ maps ordinals to predicates on terms, and is defined as follows:
\[f_p(\beta)(t)\iff\exists \gamma.\:{\beta\leq\gamma<\alpha}\wedge
{p\in\ppos{f(\gamma)}}\wedge {t\weak \subterm{f(\gamma)}{p}}\]
Thus the predicate $f_p$ is true for terms that weakly reduce
to terms which appear in principal position $p$ later in the sequence.
\begin{lem}\label{fplem}
$f_p$ is a predicate sequence.
\end{lem}
\begin{proof}
Assume $\beta\leq\gamma$ and $t\weak u$ and $f_p(\gamma)(u)$.
$f_p(\gamma)(u)$ means that there is a $\zeta\geq\gamma$ such that $p$
is a principal position in $f(\zeta)$ and $u\weak\subterm{f(\zeta)}{p}$.
In order to show that $f_p(\beta)(t)$ we can choose the same $\zeta$ as the witness;
$\beta\leq\zeta$ holds because $\beta\leq\gamma\leq\zeta$;
$t\weak\subterm{f(\zeta)}{p}$ holds because $t\weak u\weak\subterm{f(\zeta)}{p}$,
and $\weak$ is transitive.
\end{proof}
Explanation: the function $f_p$ watches the subterms in position $p$, where the ordinal parameter
can be seen as a notion of time; $f_p$ is true for a term if a weak reduct of the term will ``in the future''
appear in position $p$. The only way $f_p$ can (in time) switch from true to false is when at some point
in the reduction sequence the term in position $p$ is replaced by another that is not a reduct
of the previous.
This is only possible if a redex in a position above $p$ is contracted, i.e.\ a redex of the top layer;
in the simulated sequence this redex can be employed to replace an $r$ in position $p$
by an $l$ taken from elsewhere.

Another predicate sequence we shall be using is not dependent on the ordinal, and defined as:
$k_t(\beta)(u)\iff \neg (t\weak u)$. Thus $k_t$ is constantly true for all terms that are not weak reducts of $t$.
\begin{lem}\label{ktlem}
$k_t$ is a predicate sequence.
\end{lem}
\begin{proof}
Let $r\weak s$ and $k_t(\gamma)(s)$. Thus $\neg(t\weak s)$.
To show $k_t(\beta)(r)$ we need to show $\neg(t\weak r)$.
But $t\weak r$ gives $t\weak r\weak s$ and so by transitivity of $\weak$ the contradiction $t\weak s$.
\end{proof}

Given a rule $l\to r$ in $R$ and a predicate sequence $s$ we define
a dependent function $\xi_s:\Pi t:\Term{m}.\alpha\to\ppos{t}\to\Term{m}$ as follows:
\[
\xi_s(t)(\beta)(p) = 
\begin{cases}
l & \text{if}~ s(\beta)(\subterm{t}{p})\\
r & \text{otherwise}
\end{cases}
\]
Explanation: the reason this is a dependent function is that the type of the third parameter $\ppos{t}$ depends
on the first parameter, the term $t$. Thus $\xi_s(t)(\gamma)$ is a function of type $\ppos{t}\to\Term{m}$,
which is exactly the type of function needed to fill the holes in $\toplayer{t}$.
\begin{prop}\label{distrib}
Distributive properties of $\xi_s$ w.r.t.\ subterm selection and substitution application in top positions:
\begin{align*}
\forall p\in\tpos{t}.\:\subterm{\toplayer{t}_{\xi_s(t)(\beta)}}{p} &= \toplayer{\subterm{t}{p}}_{\xi_s(\subterm{t}{p})(\beta)}\\
\forall u\in\Terminfp{m}{\Sigma_R}.\:\toplayer{\overline{\sigma}(u)}_{\xi_s(\overline{\sigma}(u))(\beta)} &=
\overline{\sigma'}(u)\\
\mathrm{where}~\forall x.\:\sigma'(x)&=\toplayer{\sigma(x)}_{\xi_s(\sigma(x))(\beta)}
\end{align*}
\end{prop}
\begin{proof}
The first property is trivially proven by induction on the length of $p$.
The second is a corollary, because if $u\in\Terminfp{m}{\Sigma_R}$ (i.e.\ only using function symbols from the
signature of the top layer) then all variable positions
in $u$ become top positions in $\sigma(u)$.
\end{proof}

The function $\xi_s$ is used to create reducts of $t[1\searrow l]$ by reducing some of the $l$ to $r$,
guided by the predicate sequence. In particular, it is used
to replace principal subterms in a reduction sequence.
\begin{prop}\label{fundamental}
The fundamental properties of the function $\xi_s$ are:
\begin{align*}
{t\to_{RS} u} &\Rightarrow {\toplayer{t}_{\xi_s(t)(\beta)}\to_R^{=}\toplayer{u}_{\xi_s(u)(\beta)}}\\
{\beta\leq\gamma}&\Rightarrow{\toplayer{t}_{\xi_s(t)(\beta)}\weak\toplayer{t}_{\xi_s(t)(\gamma)}}
\end{align*}
\end{prop}
\begin{proof}
First property: let $q$ be the redex position. At positions $p$ parallel to $q$,
$\topequ{t}{p}{u}$ thus also $\topequ{\toplayer{t}_{\xi_s(t)(\beta)}}{p}{\toplayer{u}_{\xi_s(u)(\beta)}}$.

If $p\preceq q$ for some $p\in\ppos{t}$ then
certainly $p\in\ppos{u}$ and $\toplayer{t}_{\xi_s(t)(\beta)}$ and $\toplayer{u}_{\xi_s(u)(\beta)}$
also coincide in all positions that are prefixes of $p$.
At position $p$ itself these two terms have either $l$ or $r$. If $\xi_s(u)(\beta)(p)=r$ then
there is nothing to show, as both $l\to_R^{=}r$ and $r\to_R^{=}r$.
If $\xi_s(u)(\beta)(p)=l$ then we know by the definition of $\xi_s$ that
$s(\beta)(\subterm{u}{p})$ holds. Since we know $\subterm{t}{p}\to_{RS}\subterm{u}{p}$
we can use the predicate sequence property of $s$ to establish $s(\beta)(\subterm{t}{p})$.
This means by the definition of $\xi_s$ that $\xi_s(t)(\beta)(p)=l$, and therefore
$\toplayer{t}_{\xi_s(t)(\beta)}=\toplayer{u}_{\xi_s(u)(\beta)}$ as the terms also agree in position $p$.

Otherwise the redex $q$ is a top position,
$\subterm{t}{q}=\sigma(v)$, $\subterm{u}{q}=\sigma(w)$ for some rule $v\to w$ in $R$.
Using Proposition \ref{distrib} we get:
$\toplayer{t}_{\xi_s(t)(\beta)}=C'[\overline{\sigma'}(v)]$ and $\toplayer{u}_{\xi_s(u)(\beta)}=C'[\overline{\sigma'}(w)]$
for some context $C'$ and some substitution $\sigma'$.

So in particular: $\toplayer{t}_{\xi_s(t)(\beta)}=C'[\overline{\sigma'}(v)]\to_R C'[\overline{\sigma'}(w)]=\toplayer{u}_{\xi_s(u)(\beta)}$.

The reason for the second property is that
$s(\beta)\Leftarrow s(\gamma)$; consequently, fewer (but not more) principal subterms might be set to $l$;
again, all these can be reduced to $r$ --- this might require $\omega$ steps, if the top layer
of $t$ is infinite.
\end{proof}

Given an open reduction sequence $f:\alpha\to\Term{m}$ and a predicate sequence $s$ we write
$\toplayer{f}_s:\alpha\to\Term{m}$ for the function defined as follows (here $n$ is a finite ordinal and
$\lambda$ a limit ordinal or $0$):
\begin{align*}
\toplayer{f}_s(\lambda+2\cdot n) &= \toplayer{f(\lambda+n)}_{\xi_s(f(\lambda+n))(\lambda+n)}\\
\toplayer{f}_s(\lambda+2\cdot n+1) &= \toplayer{f(\lambda+n)}_{\xi_s(f(\lambda+n))(\lambda+n+1)}
\end{align*}
This definition may require some explanation
\begin{enumerate}[(1)]
\item Because $f$ is open, $\alpha$ is a limit ordinal and thus $\lambda+n<\alpha$ implies
$\lambda+n+n+1<\alpha$, for finite $n$. Hence $\alpha$ remains as the domain for $\toplayer{f}_s$.
\item That the case distinction into even and odd ordinals is well-defined and covers all cases
follows from the normal-form theorem for ordinals \cite[page 323]{sierpinski}.
\item
The purpose of this definition is the following.
A single reduction step $f(\beta)\to f(\beta+1)$ of the original sequence is split into two stages:
\[\toplayer{f(\beta)}_{\xi_s(f(\beta))(\beta)}\weak\toplayer{f(\beta)}_{\xi_s(f(\beta))(\beta+1)}\to_R^{=}
\toplayer{f(\beta+1)}_{\xi_s(f(\beta+1))(\beta+1)}\]
This follows from Proposition \ref{fundamental}. Thus the function $\toplayer{f}_s$ is a weak reduction sequence.
\item
As all terms in $\toplayer{f}_s$ belong to system $R$ the sequence must be convergent --- the same argument
also explains why $\toplayer{f}_s$ must be continuous at limit ordinals $\lambda<\alpha$, i.e.\ why it is a proper sequence.
\end{enumerate}
We can illustrate the construction $\toplayer{f}_s$ more concretely on
an example. 
\begin{exa}
Consider the two iTRS from Example \ref{rearrange}; they allowed to eventually rewrite $J(t)$ where
$t=K(E,t)$ to $J(u)$ where $u=K(S^\infty,u)$.
One way of doing this in $\omega+\omega$ steps is by first rewriting $t$ using rules of the first system:
$t\weak K(Z,t)\weak K(Z,K(S(z),t))\weak K(Z,K(S(Z),K(S(S(Z)),t)))\weak\ldots$.
After reaching a limit at $\omega$
we can apply the single rule of the other system repeatedly at the root:
if $u_n=K(S^n(Z),u_{n+1})$ then $J(u_n)\to J(u_{n+1})$, reaching $J(u)$ in the limit.

If this is our sequence $f$ we can form $\toplayer{f}_{f_p}$, e.g.\ for $p=1\cdot 1$.
The initial term of that sequence would be $t_0=J(a_0)$ where $a_0=K(l,a_0)$ ---
for $l$ we can choose $J(K(x,y))$ as it is a left-hand side of a non-trivial rewrite rule.
Rewriting $E$ to $Z$ in position $p$ does not change that, because all principal subterms can still rewrite to that
term, so $t_1=t_0$. Once we rewrite the other occurrences of $E$ with the rule $E\to H(E)$ they no
longer weakly reduce to $Z$, but they still weakly reduce to some term that appears
in position $p$ in the second half of sequence $f$. Therefore, $\toplayer{f}_{f_p}$ is constantly $t_0$
for the first $\omega$ steps.
The step $f(\omega)=J(u_0)\to J(u_1)=f(\omega+1)$ is split in $\toplayer{f}_{f_p}$ into
two steps: $\toplayer{f}_{f_p}(\omega)=J(a_0)\to J(K(r,a_0))\to J(a_0)=\toplayer{f}_{f_p}(\omega+2)$.
The first step rewrites the topmost $l$ to $r$, because none of the terms appearing
in position $p$ in sequence $f$ henceforth is a weak reduct of $Z$. All the other principal subterms
will still appear in $p$ and thus they will remain $l$ in $\toplayer{f}_{f_p}$.
For all subsequent steps exactly the same happens, as $S(Z)$, $S(S(Z))$, etc.\ will eventually appear
a last time in $p$. Therefore, in the second half of $\toplayer{f}_{f_p}$ the values always alternate
between $J(a_0)$ and $J(K(r,a_0))$, which makes $\toplayer{f}_{f_p}$ non-convergent.
\end{exa}

The reason a convergent reduction was mapped by this construction to a non-convergent one
is that $f$ is convergent ``in the wrong way'', i.e.\ its principal subterms in $p$ do not
form a focussed sequence, which is only possible if the system of the top layer
is itself non-convergent.


\subsection{General Observations}

\begin{lem}\label{focussed}
Let $f:\alpha\to\Term{m}$ be an open reduction sequence
and let $R$ be convergent. Let $p\in\ppos{f}$. Then the sequence $\subterm{f}{p}$ is focussed.
\end{lem}
\begin{proof}
First notice that $\toplayer{f}_{f_p}$ converges. By construction, that sequence
has an $l$ in position $p$ for all even ordinals (including limits), but it reduces that $l$ to $r$
whenever the corresponding subterm in $f$ fails to have any more reducts in position $p$.
This cannot happen arbitrarily close to $\alpha$, so there is a cut-off point $\beta$ after which
all $\subterm{f(\gamma)}{p}$ have reducts $\subterm{f(\kappa)}{p}$, for some $\kappa>\gamma$.

Now assume, for some $\zeta>\beta$ there was no cut-off point $\phi$ such that for
all $\psi\geq\phi$, $\subterm{f(\zeta)}{p}\weak\subterm{f(\psi)}{p}$.
Then we could construct the weak reduction sequence
$\toplayer{f}_{k_{\subterm{f(\zeta)}{p}}}$ in $R$.
This replaces all weak reducts of $\subterm{f(\zeta)}{p}$ in principal positions
by $r$ and all other principal
subterms by $l$. By the assumption and the previous observation we have both $l$ and $r$
occurring at $p$ arbitrarily close to $\alpha$ which contradicts convergence.
Hence the assumption was wrong and $f$ must be focussed.
\end{proof}
This result means together with the previous propositions about focussed sequences that
if reductions in principal subterm positions would converge then so would the reduction as a whole.
But why would they converge? 
We certainly have the following:
\begin{thm}\label{finiterank}
If two non-collapsing iTRSs $R$ and $S$ are convergent and $f$ is a reduction sequence of their disjoint union then
$f[n\searrow s]$ is convergent, for any $n$ and $s$.
\end{thm}
\begin{proof}
Convergence is proved by induction on $n$.

For each $p\in\ppos{f}$ the sequence $\subterm{f}{p}$ is focussed by Lemma \ref{focussed}, and so
is $\subterm{f[k+1\searrow s]}{p}=\subterm{f}{p}[k\searrow s]$.
Thus by the induction hypothesis and Proposition \ref{focussedprop} these sequences must converge.
This means we can apply Corollary \ref{divide-and-conquer}, establishing the result.
\end{proof}

\begin{exa}\label{diverge-exa}
Consider the iTRS under term metric $\infty$ with rule
$F(F(x))\to G(x)$, combined with a rule-less iTRS with function symbol $H$ and term metric $H_d(x)=\min(1,2\cdot x)$.
We can construct the infinite term $t=H(F(F(t)))$ in the combined metric.
We can also construct an open reduction sequence $g$ of length $\omega$ by starting with $t$,
always reducing its topmost redex.
By Theorem $\ref{finiterank}$ each $g[n\searrow s]$ is converging, but $g$ itself is not, the distances
between $g(k)$ and $g(k+1)$ always remain $1$.
\end{exa}
In Example \ref{diverge-exa} the sequence was diverging, because the infinite term $u=H(G(u))$ (which exists
in $\Terinf$) towards which the reduction was seemingly heading did not exist in $\Term{m}$, 
as its finite approximants do not form Cauchy-sequences under metric $d_m$. This is always the case:
\begin{prop}\label{cofinal-convergence}
If two non-collapsing iTRSs $R$ and $S$ are convergent and $f$ is a reduction sequence of their disjoint union then
$f$ converges under the metric $d_\infty$.
\end{prop}
\begin{proof}
From Theorem \ref{finiterank} we know that $f[n\searrow s]$ is converging (under metric $d_m$),
and by Proposition \ref{cofinal} we also know that it is Cauchy under metric $d_\infty$.
Under that metric the sequences $f$ and $f[n\searrow s]$ are pointwise within distance
of $2^{-n}$, and therefore $f$ must eventually stay within a ball with diameter $2^{-n}$.
As this holds for any $n$, $f$ must be Cauchy (and by completeness: converging) under metric $d_\infty$.
\end{proof}
\begin{cor}
Convergence is a modular property for non-collapsing iTRSs over metric $d_\infty$.\qed
\end{cor}
This slightly extends the result from \cite{kahrs-inf-modularity} as the proof here also covers
generalised iTRSs with extra variables.
Another consequence of Propositions \ref{cofinal-convergence} and \ref{limitexists} is that
the question whether such a sequence $f$ is converging under metric $d_m$ reduces to the question whether
its limit $t$ under $d_\infty$ is a term in $\Term{m}$, and by Proposition \ref{infinite-paths}
this is equivalent to whether all its infinite paths converge. However, infinite paths
that traverse through only finitely many signature changes would be contained in the positions of some $t[n\searrow s]$
and therefore must converge. Thus it suffices to only consider principal paths.

\subsection{Conditions on the Metric}
In general, the interplay between the umms of the constituent term metrics can be quite subtle.
As Example \ref{exa-layers2} shows, it is even possible that only this interplay gives rise
to infinite terms. In this section we are looking at metrics which are much curtailed in that respect.

A term metric $g$ is called \emph{granular} \textiff{} each component 
$\tilde{F_{i,g}}$ of the umm of each function symbol $F$
is either the identity function or the halving function. 
Clearly, when we form the disjoint union of two iTRSs with granular term metrics
the resulting term metric is granular.

As a special case $\infty$ is granular. Moreover, as indicated in the Introduction,
the term universe associated with algebraic types in Haskell
with strictness annotations is expressible through granular term metrics.

Granular term metrics have comparatively simple convergence properties: 
a full infinite path $f$ of $t\in\Terinf$ is converging w.r.t.\ term metric $g$ \textiff{}
it contains infinitely many non-strict positions. If $f$ is a path of $t$
then we define the step function as follows:
\begin{eqnarray*}
\mathrm{step}_g(t,f,0)&=&0\\
\mathrm{step}_g(t,f,n+1)&=& 
\begin{cases}
\mathrm{step}_g(t,f,n),~\text{if}~(t,f(n))_g=(t,f(n+1))_g\\
\mathrm{step}_g(t,f,n)+1,~\text{otherwise}
\end{cases}
\end{eqnarray*}
Although that definition would technically carry over for other term metrics,
for granular term metrics the step-function tells us about the rate of convergence.

Collapsing rules cause problems in term metric $\infty$ for two separate reasons:
\begin{itemize}
\item they allow to make layers merge, so that computations originating in a lower layer can impact on a higher layer;
\item the top layer can disappear altogether, so that in terms of overall convergence the computation has
made no progress up to the point where that layer collapsed.
\end{itemize}
With granular term metrics there is a different kind of rule that recreates the second problematic
aspect of collapsing rules.
This happens with a rule $l\to r$ when for some variable $x$,
$\mathit{vdepth}(x,r,g)(1)=1$ and $\mathit{vdepth}(x,l,g)(1)<1$; such rules are called \emph{pseudo-collapsing}.
Note that if all argument positions are lazy ($g=\infty$) this is only possible if $l\to r$ is a collapsing rule.

Although a pseudo-collapsing rule cannot make a layer itself disappear, it can make it disappear ``semantically'',
i.e.\ regarding distances. This has an immediate consequence on convergence:
\begin{prop}
If a non-collapsing iTRS contains a pseudo-collapsing rule then it cannot be convergent.
\end{prop}
\begin{proof}
Let $l\to r$ be that rule and $\mathit{vdepth}(x,r,g)(1)=1$ and $\mathit{vdepth}(x,l,g)(1)<1$.
We can form the infinite term $t=l[t/x]$, because $\mathit{vdepth}(x,l,g)(1)<1$.
For the rewrite step $r[t/x]\to r[r[t/x]/x]$ we notice that
$d_g(r[t/x],r[r[t/x]/x])=d_g(t,r[t/x])$ because $\mathit{vdepth}(x,r,g)(1)=1$.
This means the rewrite sequence commencing with $t$
has identical distances in each rewrite step. This can only be convergent if they are all $0$,
which means that the substitution $t/x$ must a unifier for $l$ and $r$.
This in turn is only possible if $r=x$ which is ruled out by the non-collapsing condition.
\end{proof}

As pseudo-collapsing rules break convergence we know that such rules must be absent
when we combine two non-collapsing convergent iTRSs (over term metric $g$).
Their absence also gives an impact on an invariant of the step-function:
\begin{lem}\label{step-lemma}
Let $t\weak u$ in the disjoint union of 2 non-collapsing iTRSs with joint term metric $g$.
For every principal path $q$ of $u$ there is a principal path $p$ of $t$ such that
$\forall j.\:\mathrm{step}_g(u,q,j)\geq\mathrm{step}_g(t,p,j)$.
\end{lem}
\begin{proof}
Single rewrite step:
If $q$ is a principal path not crossing the redex position then $q$ is also in $t$
and the result is immediate. Otherwise there is an $i$ such that $q(i)$ is a position above
the contractum, and $q(i+1)=q(i)\cdot v$ the position $v$ is below a variable position
of some variable $y$ in the contractum.
On the left-hand side of the step there is a path $p$ sharing the first $i$ positions with $q$,
where $p(i+1)=p(i)\cdot w$ such that $w$ is a (depth-minimal) position of $y$ in the redex.
As the applied rule is not pseudo-collapsing, if $(t,p(i))_g\neq(t,p(i+1))_g$ then we also must have
$(u,q(i))_g\neq(u,q(i+1))_g$.

The transitive closure clearly preserves the property. Transfinite reductions do not change it either
as (in the absence of rules with extra variables) limits cannot introduce new principal paths.
\end{proof}
What Lemma \ref{step-lemma} says is that the rate of convergence
cannot go down when we rewrite, provided we measure it in numbers of layers (rather than actual distances).

\begin{thm}
The disjoint union of non-collapsing iTRSs with granular term metrics preserves convergence.
\end{thm}
\begin{proof}
Let $f$ be a reduction sequence in that disjoint union.
We show that for any $\epsilon>0$ there is an $n$ such that pointwise $d_m(f(\gamma),f(\gamma)[n\searrow x])<\epsilon$;
by convergence of $f[n\searrow x]$ this implies that $f$ stays eventually with an $\epsilon$-ball, and as this holds
for any $\epsilon$, $f$ converges.

We can find the $n$ as follows: first notice that for some $k$, $\epsilon\geq 2^{-k}$.
We need to find an $n$ such that $\mathrm{step}(f(0),p,n)\geq k$ for every principal path $p$ of $f(0)$.
As $f(0)\in\Term{m}$, every principal path $p$ in it is converging which implies
$\mathrm{step}(f(0),p,n)\geq k$ for some $n$, and as $f(0)$ is a finitely branching tree,
some $n$ will do for all.

By Lemma \ref{step-lemma} this property is preserved by rewriting: 
if $t\weak u$ and $\mathrm{step}(t,p,n)\geq k$ for all principal paths $p$ of $t$ then also
$\mathrm{step}(u,q,n)\geq k$ for all principal paths $q$ of $u$. This also implies that $u$ and $u[n\searrow x]$
are within $2^{-k}$ distance of each other.
\end{proof}

\section{Modularity of strong convergence}
Simonsen showed \cite{simonsen-modularity} that \emph{top-termination} (and thus strong convergence)
is a modular property of left-linear iTRSs under term metric $\infty$.
Left-linearity is indeed an unnecessary constraint:

\begin{thm}\label{sc1}
Over term metric $\infty$, the disjoint union of two strongly converging non-collapsing iTRSs is a strongly converging iTRS.
\end{thm}
\begin{proof}
Strong convergence implies that the contributing iTRSs are non-collapsing, hence the previous
results about convergence apply.
According to \cite{zantema-inf-normalisation} an iTRS is strongly converging
if and only if it contains no reduction sequence that reduces a root redex infinitely often.
Suppose $f$ was such a sequence.
But $f[1\searrow x]$ resides in the iTRS $R$ of the top layer;
as the construction preserves all top layer reductions it also preserves
all root reductions which contradicts the premise that $R$ is strongly converging.
\end{proof}

This statement does not even need a side condition banning non-collapsing rules, since these
break the strong convergence property on the individual iTRSs and are therefore known to be absent.
For other metrics this is no longer true: term metric $\mathrm{id}$ (all components of all
umms are the identity function) rules out infinite terms and makes strong convergence
coincide with strong normalisation.

\begin{thm}
Let $R$ and $S$ be non-collapsing strongly convergent iTRSs.
If $R+S$ is convergent then it is strongly convergent.
\end{thm}
\begin{proof}
The same argument as in the proof of Theorem \ref{sc1} applies, i.e.\ if $f$ is not top-terminating
then $f[1\searrow x]$ is not either.
\end{proof}

\section{Conjectures and Open Problems}
I would conjecture that the disjoint union of a convergent iTRS with a strongly convergent one is
convergent (under term metric $\infty$). If the (merely) convergent iTRS contained no collapsing rule
then this follows already from the results in \cite{kahrs-inf-modularity}, but it may hold without that proviso.
What makes a proof for this conjecture technically awkward is that collapsing rules allow layers to merge.
However, this capability is highly restricted in converging systems.

Convergence requires that reduction sequences of \emph{any length} converge.
A related property is $\alpha$-convergence,
which means that reduction sequences up to length $\alpha$ converge, while longer ones may not.
Example \ref{ex-zantema} is not convergent in general, but it is $\omega\times n$-convergent, for finite $n$.
It depends on the ordinal $\alpha$ whether the constructions
from this paper would still work unchanged.

The most interesting special case is $\omega$-convergence, and there is a problem
with constructing the simulated reduction sequence relative to a predicate sequence:
when ``time is moved on'' the simulation potentially requires $\omega$-many steps to rewrite
potentially infinitely many occurrences of $l$ to $r$. This cannot be done if the overall length
of the simulated sequence is limited to $\omega$. Finitely many steps would suffice to move on time
if all rules were \emph{finitely linear},
which means that variables only occur a finite number of times on both left- and right-hand sides of rules.
While necessary to repair the proof it is unclear whether that constraint is necessary to
repair the theorem: are there $\omega$-convergent iTRSs that are not finitely linear and whose disjoint union
fails to be $\omega$-convergent?

\section{Conclusion}
For the standard metric $d_\infty$
it has been shown that both weak and strong convergence are modular properties of non-collapsing iTRSs.
Otherwise, the iTRSs involved can have various usually undesirable properties: non-left-linear rules,
infinite left-hand sides, and infinitely many rules are all permitted.

The proof for strong convergence is perfectly straightforward,
the proof for weak convergence is not.
At the heart of the latter is a consideration that the reductions in the top layer are limited
in the way they can rearrange the principal subterms, in particular that the principal subterms
in a fixed position must form a \emph{focussed} sequence. Moreover, abstract reduction systems
are converging \textiff{} their focussed sequences do, and both observations together give that result.

For other metrics, these arguments still apply to reductions of finite rank (Theorem \ref{finiterank}), while
for reductions of infinite rank there are counterexamples.
As a general proof strategy, various convergence arguments have been constructed along the central idea
of limiting the pointwise distance between the reduction sequence under investigation
and a finite rank approximation of that sequence. These allow to generalise the modularity of convergence
and strong convergence of non-collapsing iTRS beyond metric $d_\infty$,
to systems with either granular term metrics or (not elaborated here)
arbitrary metrics but depth-preserving rules.
\bibliographystyle{plain}
\bibliography{lambda,trs,rewrite,trees,set,general,func}

\end{document}